\documentclass[twocolumn,amsmath,amssymb]{revtex4}

\usepackage{graphicx}
\usepackage{dcolumn}
\usepackage{bm}

\usepackage{epsf}
\usepackage{subfigure}
\usepackage{epstopdf}
\begin{document}

\title{Exactly solvable model of quantum diffusion}

\author{M. Esposito}
\author{P. Gaspard}%
\affiliation{Center for Nonlinear Phenomena and Complex Systems,\\
Universit\'e Libre de Bruxelles, Code Postal 231, Campus Plaine, B-1050
Brussels, Belgium.}

\date{\today}

\begin{abstract}
We study the transport property of diffusion in a finite translationally 
invariant quantum subsystem described by a tight-binding Hamiltonian with
a single energy band and interacting with its environment by a coupling in 
terms of correlation functions which are delta-correlated in space and time.
For weak coupling, the time evolution of the subsystem density matrix
is ruled by a quantum master equation of Lindblad type. 
Thanks to the invariance under spatial translations, we can apply the 
Bloch theorem to the subsystem density matrix and exactly diagonalize 
the time evolution superoperator to obtain the complete spectrum of its 
eigenvalues, which fully describe the relaxation to equilibrium.
Above a critical coupling which is inversely proportional to the size of the subsystem, 
the spectrum at given wavenumber contains an isolated eigenvalue describing diffusion. 
The other eigenvalues rule the decay of the populations and quantum coherences
with decay rates which are proportional to the intensity of the environmental noise. 
On the other hand, an analytical expression is obtained for the dispersion
relation of diffusion.  The diffusion coefficient is proportional to 
the square of the width of the energy band and inversely proportional 
to the intensity of the environmental noise because diffusion results 
from the perturbation of quantum tunneling by the environmental
fluctuations in this model. Diffusion disappears below the critical coupling.
\end{abstract}


\keywords{Transport processes, quantum diffusion, transport in
mesoscopic systems, decoherence, Redfield equation, Liouvillian resonance.}

\maketitle
\section{Introduction \label{intro}}

The diffusion of particles in a condensed phase is a fundamental
transport process which is ubiquitous
in natural phenomena. Since the pioneering work by Einstein in 1905,
it is known that diffusion is related to conduction or mobility.
Although diffusion has been extensively studied in classical systems,
much less is known in
quantum systems where the quantum effects can deeply affect the
transport properties at
low temperature. 
Yet, the theoretical understanding of quantum diffusion remains sparse 
and difficult because diffusion is an incoherent
process very remote from the very coherent basic quantum dynamics.

The purpose of the present paper is to study a simple
model of quantum diffusion
in a system which possesses the property of being invariant under spatial
translations. This property is a fundamental feature of systems
sustaining transport processes
such as diffusion or conduction. The translational invariance is at the basis
of the early studies of electronic conduction based on the
Boltzmann-Lorentz equation \cite{Ashcroft},
and is also important in the polaron model \cite{Holstein}. Motivated
by the problem of dissipation
in quantum macroscopic phenomena, models of quantum Brownian motion have been
proposed and studied since the eighties \cite{Caldeira}. In such
models, one degree of freedom
has a diffusive-like motion in an external potential which breaks the
translational invariance.
These models were later extended to the study of transport in
spatially periodic potentials 
\cite{Schmid,Fisher,Weiss1,Weiss2,Weiss,Egger,Lebowitz1,Lebowitz2,Lebowitz3}.
In these studies, attention was focused essentially on diffusion.
In the present study, we intend to study the complete set of the relaxation 
modes in the system by obtaining all the eigenvalues of the time evolution 
superoperator. 
The complete diagonalization of the time evolution superoperator 
is accomplished by fully exploiting the translational invariance 
with Bloch's theorem applied to the density matrices.  
This allows us to introduce in a rigorous way a wavenumber and to 
obtain the eigenvalues which gives the Liouvillian resonances as 
functions of the wavenumber.
These resonances provides us with the characteristic times
of the relaxation toward the thermodynamic equilibrium.
The hydrodynamic mode of diffusion is one among
the eigenstates associated with the Liouvillian resonances.
This mode controls the long-time dynamics of the system.
The other modes control the shorter time scales 
and are associated with decoherence. 
In the quantum model here studied, analytical expressions can be obtained 
in this way for the dispersion relation of diffusion and the other resonances.

We notice that translationally invariant models without coupling to
the environment
have recently been studied in such models as quantum graphs \cite{Barra}
quantum multibaker maps \cite{Dorfman}, as well as quantum periodic
Lorentz gases \cite{Knauf}.
These models describe the motion of a quantum particle in a spatially
periodic potential.
The energy spectrum of the particle is composed of energy bands so
that the motion
is ballistic on long-time scales. If the energy of the particle is
distributed over many
energy bands, the motion becomes semiclassical and diffusion may
manifest itself
as a transient behavior before the long-time ballistic motion. This has been
remarkably demonstrated for the quantum multibaker map in Ref. \cite{Dorfman}.
In order for diffusion to persist up to arbitrarily long times, the
quantum system must
contain infinitely many degrees of freedom, for instance, by the
coupling with some
environment in the subsystem-plus-reservoir approach we adopt in the
present paper.

The plan of the paper is the following.
Our translationally invariant model is defined in Sec. \ref{Sec.system} 
by weakly coupling a one-dimensional tight-binding Hamiltonian
to a delta-correlated environment. 
The dynamics of this system is ruled by a Redfield quantum master equation
\cite{Red,Kampen,KuboB2,Breuer,GaspRed}.
The long-time evolution of the model can be studied in terms of the eigenvalues
and the associated eigenstates of the Redfield superoperator, as explained in
Sec. \ref{Sec.diag}. 
The eigenvalue problem is exactly solved for a finite chain in Sec. \ref{Sec.finite}. 
The advantage of first taking a finite chain is that the total number of eigenvalues is known
and we do not miss the part of the spectrum controlling decoherence.
The Liouvillian spectrum of the infinite chain is thereafter obtained in Sec. \ref{Sec.infinite}
by taking the limit $N\to\infty$. 
Conclusions are drawn in Sec. \ref{Sec.conclusions}.


\section{Defining the system \label{Sec.system}}

\subsection{Subsystem}

We consider a one-dimensional subsystem
described by the following Hamiltonian
\begin{eqnarray}
\hat{H}_{\rm s}=
\left(\begin{array}{ccccccc}
E_0 & -A & 0 & 0 & \hdots & 0 & -A \\
-A & E_0 & -A & 0 & \hdots & 0 & 0 \\
0 & -A & E_0 & -A & & 0 & 0 \\
\vdots & &\ddots &\ddots & \ddots & & \vdots \\
0 & 0 & & -A & E_0 & -A & 0 \\
0 & 0 & \hdots & 0 & -A & E_0 & -A \\
-A & 0 & \hdots & 0 & 0 & -A & E_0 \\
\end{array} \right)_{N \times N} \nonumber\\ \label{isolaaa}
\end{eqnarray}
represented in the site basis ${\vert l \rangle}$, where $l$ takes
the values $l=0,1,\hdots,N-1$. $N$ is the length of the chain.
We have here chosen periodic (Born-van Karman) boundary conditions.
This so-called tight-binding or H\"uckel Hamiltonian 
describes a process of quantum tunneling from site to site
and  is invariant under spatial translations. 

The stationary Schr\"{o}dinger equation of the tight-binding
Hamiltonian is given by
\begin{eqnarray}
\hat{H}_{\rm s} \vert k \rangle = \epsilon_k \vert k \rangle ,\label{isolaab}
\end{eqnarray}
where the eigenvalues are
\begin{eqnarray}
\epsilon_k=E_0-2A \cos k \frac{2 \pi}{N} \label{isolaac}
\end{eqnarray}
and the eigenvectors
\begin{eqnarray}
\langle l \vert k \rangle = \frac{1}{\sqrt{N}} {\rm e}^{i l k \frac{2
\pi}{N}} , \label{isolaad}
\end{eqnarray}
with $k=0,1,...,N-1$.
Accordingly, the Hamiltonian (\ref{isolaaa}) has an energy spectrum with
a single energy band of width $4A$ and the motion of the particle
would be purely ballistic without coupling to a fluctuating environment.

\subsection{Coupling to the environment \label{coupling}}

Now, we suppose that the subsystem is coupled to a large environment.
The Hamiltonian of the total system composed of the one-dimensional
chain and its environment is
given by
\begin{eqnarray}
\hat{H}_{\rm tot} = \hat{H}_{\rm s} + \hat{H}_{\rm b} + \lambda \sum_{l}
\hat{S}_{l} \hat{B}_{l} \; ,\label{isolaaf}
\end{eqnarray}
where $\hat{H}_{\rm b}$ is the environment Hamiltonian,
$\hat{S}_{l}$ the subsystem coupling operators,
$\hat{B}_{l}$ the environment coupling operators, and $\lambda$
the coupling parameter which measures the intensity of the
interaction between the subsystem and its environment.

The dynamics of the total system is described by the von Neumann equation
\begin{eqnarray}
\frac{d \hat{\rho}_{\rm tot}(t)}{dt}=\hat{\hat{{\cal L}}}_{\rm tot}
\hat{\rho}_{\rm tot}(t)
=-i [\hat{H}_{\rm tot},\hat{\rho}_{\rm tot}(t)], \label{isolabf}
\end{eqnarray}
where $\hat{\hat{{\cal L}}}_{\rm tot}$ is the Liouvillian
superoperator of the total system.
We adopt the convention that $\hbar=1$.
The reduced dynamics for the density matrix $\hat{\rho}(t)={\rm
Tr}_{\rm b} \hat{\rho}_{\rm tot}(t)$ of the subsystem is known to
obey a Redfield quantum master equation for weak coupling to the 
environment \cite{Red,Kampen,KuboB2,Breuer,GaspRed}.
This equation can be systematically derived from the complete von
Neumann equation for the total system (\ref{isolabf}) by second-order
perturbation theory so that
\begin{eqnarray}
\frac{d \hat{\rho}(t)}{dt} = {\rm Tr}_{\rm b} \hat{\hat{{\cal
L}}}_{\rm tot} \hat{\rho}_{\rm tot}(t)
\stackrel{\lambda \ll 1}{=} \hat{\hat{{\cal L}}}_{\rm Red}(t) \hat{\rho}(t)
+ O(\lambda^3) .\label{isolacf}
\end{eqnarray}
On time scales longer than the correlation time of the environment,
the Redfield quantum master equation is Markovian and reads
\begin{eqnarray}
\frac{d \hat{\rho}}{dt} &=& \hat{\hat{{\cal L}}}_{\rm Red} \hat{\rho}
\nonumber \\
&=& -i \lbrack \hat{H}_{\rm s} , \hat{\rho} \rbrack + \lambda^2 \sum_{l} (
\hat{T}_{l} \hat{\rho} \hat{S}_{l} \nonumber \\ &+&
\hat{S}_{l}^{\dagger} \hat{\rho}
\hat{T}_{l}^{\dagger} - \hat{S}_{l} \hat{T}_{l}
\hat{\rho} - \hat{\rho} \hat{T}_{l}^{\dagger}
\hat{S}_{l}^{\dagger} ) + O(\lambda^3), \label{isolaag}
\end{eqnarray}
where $\hat{\hat{{\cal L}}}_{\rm Red}$ is the so-called Redfield
superoperator and where
\begin{eqnarray}
\hat{T}_{l} \equiv \sum_{l'} \int_{0}^{\infty} d \tau \;
C_{l l'}(\tau) \; {\rm e}^{-{i} \hat{H}_{\rm s} \tau}
\hat{S}_{l'} \;
{\rm e}^{i \hat{H}_{\rm s} \tau} . \label{isolaah}
\end{eqnarray}
The correlation function of the environment which contains all the
necessary information to describe the coupling of the subsystem to
its environment is given by
\begin{eqnarray}
C_{l l'} (\tau) = {\rm Tr}_{\rm b} \hat{\rho}_{\rm b}^{\rm
eq} {\rm e}^{i \hat{H}_{\rm b} \tau} \hat{B}_{l} {\rm e}^{-i
\hat{H}_{\rm b} \tau} \hat{B}_{l'} \label{isolaai}
\end{eqnarray}
where $\hat\rho_{\rm b}^{\rm eq}$ is the canonical equilibrium state
of the environment.

The interaction of the subsystem with its environment is expressed
in terms of the subsystem coupling operators which are projection
operators on the site basis
\begin{eqnarray}
\langle l \vert \hat{S}_{l''} \vert l' \rangle= \delta_{l l'} \; \delta_{l l''} \; .
\label{isolaal}
\end{eqnarray}
They take the unit value if the particle is located on the site 
$l''$ and zero otherwise.
These operators have the properties:
\begin{eqnarray}
\hat{S}_{l}^n = \hat{S}_{l} \; ,
\end{eqnarray}
for $n=2,3,...$ and
\begin{eqnarray}
\sum_{l} \hat{S}_{l} = \hat{I} \; .
\end{eqnarray}

We now need to specify the environment operators
by the choice of the environment correlation functions.
We make two assumptions:\\
\underline{Assumption 1}: The environment dynamics has a very fast
evolution compared with the subsystem dynamics. Therefore, the
environment correlation functions decay in time so fast
with respect to the characteristic time scales of the subsystem evolution,
that they can be assumed to be Dirac delta distributions in time.\\
\underline{Assumption 2}: The environment has very short range
spatial correlations,
much shorter than the distance between two adjacent sites of the subsystem.
Accordingly, the environment correlation functions decay in space
so fast that they can be assumed to be given by Kronecker delta in space.\\
These two assumptions mean that each of the environmental
fluctuations at the different subsystem sites are statistically
independent. The environment correlation functions are thus given by
\begin{eqnarray}
C_{l l'}(\tau)= 2\; Q \; \delta (\tau)\; \delta_{l l'} \; , \label{isolaaj}
\end{eqnarray}
where $Q$ is a real number.  The operators (\ref{isolaah}) in the 
Redfield equation (\ref{isolaag}) therefore become
\begin{eqnarray}
\hat{T}_{l} = Q \; \hat{S}_{l} \; . \label{isolaak}
\end{eqnarray}
It has been shown in Ref. \cite{Esposito} that the form taken 
by the operator (\ref{isolaak}) can be physically justified when
$\tau_{\rm th} \ll \tau_{\rm b} \ll \tau_{\rm s}$, where $\tau_{\rm th}=1/k_{\rm B}T$
is the thermal time, $\tau_{\rm b}$ the correlation time of the environment or bath,
and $\tau_{\rm s}$ the subsystem characteristic time.

Because of the fast decay of the temporal and the spatial
correlations (\ref{isolaaj})
and the properties of the subsystem coupling operators
(\ref{isolaal}), the Redfield equation we have to solve takes the form
\begin{eqnarray}
\frac{d \hat{\rho}}{dt}
&=& \hat{\hat{{\cal L}}}_{\rm Red} \hat{\rho} \label{isolaam} \\
&=& -i \lbrack \hat{H}_{\rm s} ,
\hat{\rho} \rbrack \nonumber \\ && + \lambda^2 Q \sum_{l} (
2 \hat{S}_{l} \hat{\rho} \hat{S}_{l}
- \hat{S}_{l}^2 \hat{\rho} - \hat{\rho} \hat{S}_{l}^2)
+ O(\lambda^3) \; . \nonumber
\end{eqnarray}
It can easily be verified by projecting this equation onto the site
basis that it is translationally invariant
(shifting all the site indices appearing in the projected equation by
a constant does not modify the equation).
Furthermore, this equation preserves the complete positivity of the
density matrix because it has the Lindblad form \cite{Lindblad} which
is the result of a coupling with delta correlation functions
\cite{GaspRed}.
It should also be mentioned that Eq. (\ref{isolaam}) can be directly derived 
from the complete von Neumann equation for the total system (\ref{isolabf})
in the singular coupling limit \cite{Gorini,Spohn80}. 


\section{Diagonalizing the Redfield superoperator \label{Sec.diag}}

The eigenvalues $s_{\nu}$ and associated eigenstates
${\hat{\rho}}^{\nu}$ of the Redfield superoperator are defined by
\begin{eqnarray}
\hat{\hat{{\cal L}}}_{\rm Red}\; {\hat{\rho}}^{\nu} = s_{\nu} \;
{\hat{\rho}}^{\nu} , \label{blochaaa}
\end{eqnarray}
where $\nu$ is a set of parameters labeling the eigenstates. Because the
Redfield superoperator is not anti-Hermitian, its eigenvalues can be
complex numbers
with a nonzero real part. The eigenvalue problem of the Redfield
superoperator is important because
the time evolution of the quantum master equation can then be
decomposed onto the basis of the eigenstates as
\begin{eqnarray}
\hat{\rho}(t) = {\rm e}^{\hat{\hat{{\cal L}}}_{\rm Red} t} \hat{\rho}(0)
= \sum_{\nu=1}^{N^2} c_{\nu}(0) \; {\rm e}^{s_{\nu} t} \hat{\rho}^{\nu}.
\label{isolaan}
\end{eqnarray}
The dynamics is therefore given by a linear superposition of
exponential or oscillatory exponential
functions. Since the reduced density matrix of the subsystem has
$N^2$ elements,
there is a total of $N^2$ eigenvalues and associated eigenstates.

\subsection{Bloch theorem \label{Bloch}}

Since the system is invariant under spatial translations, we can
apply the Bloch theorem to
the eigenstates of the Redfield superoperator. Thanks to this
theorem, the state space of the
superoperator can be decomposed into independent superoperators
acting onto decoupled sectors associated with a
given Bloch number, also called wavenumber \cite{Ashcroft}.

We define the superoperator $\hat{\hat{{\cal T}}}_a$ of the
spatial translation by $a$ sites along the system ($a$ is an integer)
as
\begin{eqnarray}
(\hat{\hat{{\cal T}}}_a \rho^{{\nu}})_{ll'}
= \rho^{{\nu}}_{l+a,l'+a} \; , \label{blochaab}
\end{eqnarray}
where we use the notation
$\langle l \vert \hat{\rho} \vert l' \rangle = \rho_{ll'}$.
This superoperator has the group property
\begin{eqnarray}
\hat{\hat{{\cal T}}}_a \hat{\hat{{\cal T}}}_{a'}
= \hat{\hat{{\cal T}}}_{a'} \hat{\hat{{\cal T}}}_a = \hat{\hat{{\cal
T}}}_{a+a'} \; . \label{blochaac}
\end{eqnarray}
Because of the translational symmetry of the system, the translation
superoperators commute
with the Redfield superoperator
\begin{eqnarray}
\left[ \hat{\hat{{\cal T}}}_a , \hat{\hat{{\cal L}}}_{\rm Red} \right] =0
 \; . \label{blochaad}
\end{eqnarray}
Therefore, the Redfield superoperator as well as the translation
superoperators have a basis of
common eigenstates.
If $\tau(a)$ denotes the eigenvalues of the translation
superoperator, we have that
\begin{eqnarray}
\hat{\hat{{\cal T}}}_a \; {\hat{\rho}}^{{\nu}} = \tau(a) \;
{\hat{\rho}}^{{\nu}} \; , \label{blochaae}
\end{eqnarray}
where, because of the unitarity of $\hat{\hat{{\cal T}}}_a$,
\begin{eqnarray}
\vert \tau(a) \vert^{2}=1 \; . \label{blochaaf}
\end{eqnarray}
Equation (\ref{blochaac}) implies
\begin{eqnarray}
\tau(a+a')=\tau(a) \; \tau(a') \; . \label{blochaag}
\end{eqnarray}
Because of Eqs. (\ref{blochaaf}) and (\ref{blochaag}), we find that
\begin{eqnarray}
\tau(a)={\rm e}^{i q a} \; ,\label{blochaah}
\end{eqnarray}
where $q$ is the Bloch number or wavenumber, 
whereupon we get
\begin{eqnarray}
\rho^{{\nu}}_{l+a,l'+a} = {\rm e}^{iqa} \rho^{{\nu}}_{ll'} \; .
\label{blochaai}
\end{eqnarray}
A useful consequence is that
\begin{eqnarray}
\rho^{{\nu}}_{ll'} = {\rm e}^{iql} \rho^{{\nu}}_{0,l'-l} \; .\label{blochaaj}
\end{eqnarray}
In order to determine the allowed values of the Bloch number, we
write by using Eq. (\ref{isolaad}) that
\begin{eqnarray}
\rho^{\nu}_{ll'} = \frac{1}{N} \sum_{k,k'} \langle k \vert
\hat{\rho}^{\nu} \vert k' \rangle
{\rm e}^{i (l k - l' k') \frac{2 \pi}{N}} \label{blochaak}
\end{eqnarray}
and
\begin{eqnarray}
\rho^{\nu}_{l+1,l'+1} &=& \frac{1}{N} \sum_{k,k'} \langle k \vert
\hat{\rho}^{\nu} \vert k' \rangle
{\rm e}^{i (l k - l' k') \frac{2 \pi}{N}} {\rm e}^{i (k - k') \frac{2
\pi}{N}} . \label{blochaal}
\end{eqnarray}
Because of Eq. (\ref{blochaai}), we also have
\begin{eqnarray}
\rho^{\nu}_{l+1,l'+1} &=& {\rm e}^{iq} \; \rho^{\nu}_{ll'} \; .\label{blochabl}
\end{eqnarray}
Multiplying both sides of Eqs. (\ref{blochaal}) and (\ref{blochabl}) by
$\langle l' \vert k''' \rangle \langle k'' \vert l \rangle$, taking the sum
$\sum_{l,l'}$ of it, and identifying them, we obtain
\begin{eqnarray}
{\rm e}^{iq} \langle k \vert \hat{\rho}^{\nu} \vert k' \rangle = {\rm
e}^{i (k - k') \frac{2 \pi}{N}}
\langle k \vert \hat{\rho}^{\nu} \vert k' \rangle. \label{blochaam}
\end{eqnarray}
We can now notice that if $q \neq (k-k')\frac{2 \pi}{N}$, then
$\langle k \vert \hat{\rho}^{\nu} \vert k' \rangle=0$.
Finally, using the periodicity
\begin{eqnarray}
\rho_{l+N,l'}^{{\nu}}&=&\rho_{ll'}^{{\nu}} \; , \\
\rho_{l,l'+N}^{{\nu}}&=&\rho_{ll'}^{{\nu}} \; , \label{blochaao}
\end{eqnarray}
and Eq. (\ref{blochaai}), we find that the Bloch number takes the values
\begin{eqnarray}
q = j \frac{2 \pi}{N}, \ \ \text{where} \ \ j=0,1,\hdots,N-1 \; .
\label{blochabo}
\end{eqnarray}
Consequently, the Redfield superoperator can be block-diagonalized into $N$
independent blocks, which each contains $N$ eigenvalues as we shall
see in the following.

\subsection{Simplifying the problem \label{diag}}

The eigenvalue problem of the Redfield superoperator
can be formulated in each sector labeled by a given wavenumber $q$.
For this purpose, Eq. (\ref{blochaaa}) with the explicit expression
(\ref{isolaam}) of the Redfield superoperator is projected onto the site basis to get
\begin{eqnarray}
s_{\nu}\rho^{\nu}_{ll'}&=& - i A \left(- \rho^{\nu}_{l-1,l'}
- \rho^{\nu}_{l+1,l'} + \rho^{\nu}_{l,l'-1} +
\rho^{\nu}_{l,l'+1}\right) \nonumber \\
&&+ 2 \lambda^2 Q \left(\delta_{ll'} - 1\right) \rho^{\nu}_{ll'} \; .
\label{spec1aaa}
\end{eqnarray}
Using Eq. (\ref{blochaaj}) and replacing $l'-l$ by $l$, we have
\begin{eqnarray}
\left(s_{\nu} + 2 Q \lambda^2\right) \rho^{{\nu}}_{0l} &=&
2 A \left(\sin\frac{q}{2}\right) \left( {\rm e}^{-i \frac{q}{2}}
\rho^{{\nu}}_{0,l+1} -
{\rm e}^{i \frac{q}{2}} \rho^{{\nu}}_{0,l-1} \right)
\nonumber \\ & & + 2 Q \lambda^2 \rho^{{\nu}}_{00} \; \delta_{0l} \; .
\label{spec1aac}
\end{eqnarray}
Making the change of variable
\begin{eqnarray}
\rho^{{\nu}}_{0l}=i^{-l} {\rm e}^{i \frac{q}{2} l} f_l \; ,
\end{eqnarray}
we obtain the simpler eigenvalue equation
\begin{eqnarray}
\mu_{\nu} f_l &=& \delta_{0l} f_l - i \beta (f_{l-1}+f_{l+1}) \;
,\label{spec1aad}
\label{eqvalproprefinale}
\end{eqnarray}
where
\begin{eqnarray}
\mu_{\nu}=\frac{s_{\nu}}{2Q \lambda^2} +1 \; ,\label{spec1baa}
\end{eqnarray}
and
\begin{eqnarray}
\beta=\frac{A}{Q \lambda^2} \sin\frac{q}{2} \; . \label{spec1bab}
\end{eqnarray}


\section{Finite chain \label{Sec.finite}}

\subsection{The eigenvalue problem}

The expression (\ref{eqvalproprefinale}) can be written in matrix form
without the index ${\nu}$ to simplify the notation,
\begin{eqnarray}
\mu \vec{f} = \hat{W} \vec{f} \; , \label{spec1abe}
\end{eqnarray}
where $\mu$ denotes the eigenvalue, $\vec{f}=(f_{0},\hdots,f_{N-1})$ the
eigenvector of size
$N$, and $\hat{W}$ the following $N \times N$ matrix
\begin{eqnarray}
\left(\begin{array}{ccccccccc}
1 & -i \beta & &
& & & -i \beta i^{-N} {\rm e}^{i N \frac{q}{2}} \\
-i \beta & 0 & -i \beta &
& & & \\
& -i \beta & 0 & -i \beta
& & & \\
& & \ddots & \ddots
& \ddots & & \\
& & & -i \beta
& 0 & -i \beta & \\
& & &
& -i \beta & 0 & -i \beta \\
-i \beta i^{N} {\rm e}^{-i N \frac{q}{2}} & & &
& & -i \beta & 0 \\
\end{array} \right) \nonumber\\ \label{spec1aae}
\end{eqnarray}
We look for eigenstates of the form
\begin{eqnarray}
f_{l} = A {\rm e}^{i \theta l} + B {\rm e}^{-i \theta l}.\label{spec1aaf}
\end{eqnarray}
Solving Eq. (\ref{spec1abe}) with (\ref{spec1aaf}) gives
\begin{itemize}
\item for $0 < l < N-1$:
\begin{eqnarray}
\mu = -2i \; \beta \; \cos \theta \; , \label{spec1aag}
\end{eqnarray}
\item for $l=0$:
\begin{eqnarray}
&& (1-\mu)(A+B) - i \beta (A {\rm e}^{i \theta}+B {\rm e}^{-i
\theta}) \nonumber \\
&& - i \beta i^{-N} {\rm e}^{i N \frac{q}{2}} (A {\rm e}^{i \theta
(N-1)}+B {\rm e}^{-i \theta (N-1)}) =0 \; ,\label{spec1aah}
\end{eqnarray}
\item for $l=N-1$:
\begin{eqnarray}
A {\rm e}^{i \theta N}+B {\rm e}^{-i \theta N} - i^N {\rm e}^{-i
N\frac{q}{2}} (A+B) =0 \; .\label{spec1aai}
\end{eqnarray}
\end{itemize}
Solving the homogeneous linear system of Eqs. (\ref{spec1aah}) and
(\ref{spec1aai}) and replacing
$\mu$ by (\ref{spec1aag}), one gets the characteristic equation
\begin{eqnarray}
2 i \beta \sin \theta \left[ R(q) - \cos \theta N \right] = \sin
\theta N, \label{spec1aaj}
\end{eqnarray}
with
\begin{eqnarray}
R(q)=\frac{1}{2}\left(i^{N} {\rm e}^{-i N \frac{q}{2}} + i^{-N} {\rm
e}^{i N \frac{q}{2}} \right). \label{spec1aatheta}
\end{eqnarray}
Using Eq. (\ref{blochabo}), we find:
\begin{eqnarray}
&\mbox{for}& \; N \; \mbox{odd}: \qquad R(q_j)=0 \; , \\
&\mbox{for}& \; N=4I: \qquad R(q_j)=(-1)^j \; ,\\
&\mbox{for}& \; N=4I+2: \qquad R(q_j)=-(-1)^j \; ,
\end{eqnarray}
with $I$ integer.
From now on, we shall speak of even (respectively odd) $q$, if
$q$ corresponds to an even (respectively odd) integer $j$ in Eq.
(\ref{blochabo}).
Therefore, the characteristic equation (\ref{spec1aaj}) becomes
\begin{itemize}
\item for $N$ odd:
\begin{eqnarray}
2 i \beta \sin \theta = - \tan \theta N \; ; \label{spec1aal}
\end{eqnarray}
\item for $N=4I$ and $q$ even or for $N=4I+2$ and $q$ odd:\\
either
\begin{eqnarray}
2 i \beta \sin \theta &=& {\rm cotan} \frac{\theta N}{2} \; , \label{spec1aam}
\end{eqnarray}
or
\begin{eqnarray}
\cos \theta N &=& 1 \; ; \label{spec1abm}
\end{eqnarray}
\item for $N=4I$ and $q$ odd or for $N=4I+2$ and $q$ even:\\
either
\begin{eqnarray}
2 i \beta \sin \theta &=& - \tan \frac{\theta N}{2} \; , \label{spec1aan}
\end{eqnarray}
or
\begin{eqnarray}
\cos \theta N &=& -1 \; .\label{spec1abn}
\end{eqnarray}
\end{itemize}
We solve Eqs. (\ref{spec1aal})-(\ref{spec1abn}) as follows.

\subsection{The diffusive eigenvalue $\mu^{(1)}$}

We first look for an eigenvalue $\mu$ which is real and should
correspond to a monotonic
exponential decay. With this goal, we suppose that the angle
$\theta$ is complex
\begin{eqnarray}
\theta = \xi + i \eta \; , \label{theta}
\end{eqnarray}
so that the eigenvalue (\ref{spec1aag}) becomes
\begin{eqnarray}
\mu = -2i \beta \cos \theta = - 2 \beta \sin\xi \sinh\eta - 2i\beta
\cos\xi\cosh\eta \; .
\end{eqnarray}
This eigenvalue is real under the condition that $\cos\xi=0$
which is satisfied for
\begin{eqnarray}
\xi = - \frac{\pi}{2} \; , \label{xi}
\end{eqnarray}
in which case
\begin{eqnarray}
\mu = - 2 \; \beta \; \sinh\eta \; . \label{cmu}
\end{eqnarray}
We notice that the condition $\cos\xi=0$ is also satisfied for
$\xi=\frac{\pi}{2}$
but it can be shown that this other case leads to the same eigenvalue
as (\ref{xi}).
If we introduce the conditions (\ref{theta}) and (\ref{xi}) in Eqs.
(\ref{spec1aal}),
(\ref{spec1aam}), and (\ref{spec1aan}), we get
\begin{itemize}
\item for $N$ odd:
\begin{eqnarray}
2 \beta \cosh \eta = \coth N\eta \; ; \label{cspec1aal}
\end{eqnarray}
\item for $N=4I$ and $q$ even or for $N=4I+2$ and $q$ odd:\\
\begin{eqnarray}
2 \beta \cosh \eta &=& \coth \frac{N\eta}{2} \; ; \label{cspec1aam}
\end{eqnarray}
\item for $N=4I$ and $q$ odd or for $N=4I+2$ and $q$ even:\\
\begin{eqnarray}
2 \beta \cosh \eta &=& \tanh \frac{N\eta}{2} \; . \label{cspec1aan}
\end{eqnarray}
\end{itemize}

In the limit $N\to\infty$, the right-hand side of these equations
tends to unity if
a nonvanishing solution $\eta \neq 0$ exists. In this limit, this solution
is thus given by
\begin{eqnarray}
\eta_0 &=& {\rm arccosh} \; \frac{1}{2 \beta} \; ,\label{spec2abp}
\end{eqnarray}
which exists only if $\beta \leq \frac{1}{2}$.
Because
\begin{eqnarray}
\sinh \eta_0 &=& \sqrt{\left(\frac{1}{2 \beta}\right)^2 - 1} \;
,\label{spec2aar}
\end{eqnarray}
the corresponding eigenvalue should be
\begin{eqnarray}
\mu_0 &=& \sqrt{1-(2 \beta)^2} \; .\label{spec2aas}
\end{eqnarray}

However, for a finite chain with $N<\infty$, we expect a correction
$\delta\eta$ to the solution
$\eta=\eta_0+\delta\eta$. Replacing this correction in Eqs. (\ref{cspec1aal}),
(\ref{cspec1aam}), and (\ref{cspec1aan}), we obtain by Taylor expansion that
\begin{itemize}
\item for $N$ odd:
\begin{eqnarray}
\delta \eta \simeq 2 \; \frac{{\rm e}^{- 2 N {\rm arccosh} \;
\frac{1}{2 \beta}}}{\sqrt{1-(2 \beta)^2}} \; ; \label{spec3abc}
\end{eqnarray}
\item for $N=4I$ and $q$ even or for $N=4I+2$ and $q$ odd:
\begin{eqnarray}
\delta \eta \simeq 2 \; \frac{{\rm e}^{- N {\rm arccosh} \;
\frac{1}{2 \beta}}}{\sqrt{1-(2 \beta)^2}} \; ;
\label{spec3acc}
\end{eqnarray}
\item for $N=4I$ and $q$ odd or for $N=4I+2$ and $q$ even:
\begin{eqnarray}
\delta \eta \simeq - 2 \; \frac{{\rm e}^{- N {\rm arccosh} \;
\frac{1}{2 \beta}}}{\sqrt{1-(2 \beta)^2}} \; ;
\label{spec3adc}
\end{eqnarray}
\end{itemize}
up to corrections of $O(\delta\eta^2)$.
Using the expression (\ref{cmu}), we finally obtain the eigenvalue
\begin{eqnarray}
\mu^{(1)} &=& 2\beta \sinh \eta_0 + 2 \beta \cosh \eta_0 \; \delta
\eta + O(\delta \eta^2) \nonumber \\
&=& \sqrt{1-(2 \beta)^2} + \delta \eta + O(\delta\eta^2)\; ,\label{spec3aad}
\end{eqnarray}
where $\delta \eta$ is respectively given by Eqs. (\ref{spec3abc}),
(\ref{spec3acc}), and (\ref{spec3adc}).
Accordingly, the correction $\delta\eta$ to the eigenvalue decreases
exponentially fast with the
size $N$ of the chain.

Using Eqs. (\ref{spec1baa}) and (\ref{spec1bab}), we finally obtain
the eigenvalue
\begin{eqnarray}
s^{(1)} &=& 2 Q \lambda^2 (\mu^{(1)} - 1) \nonumber\\
&=& 2 Q \lambda^2 \sqrt{1 - \left( \frac{2A}{Q\lambda^2} \sin
\frac{q}{2}\right)^2} -2Q\lambda^2 + O(\delta\eta) \; . \nonumber\\
\label{spec3aae}
\end{eqnarray}
The eigenvalue $s^{(1)}=0$ corresponding to a vanishing wavenumber
$q=0$ is always in the spectrum
of the Redfield superoperator. The associated eigenstate describes
the stationary equilibrium state.
At low wavenumbers $q,\beta\to 0$, we recover the dispersion relation
of diffusion
\begin{eqnarray}
s^{(1)} = -D q^2 + O(q^4)\; , \label{spec3aaf}
\end{eqnarray}
with the diffusion coefficient
\begin{eqnarray}
D = \frac{A^2}{Q \lambda^2} \; , \label{spec3aag}
\end{eqnarray}
which justifies calling $\mu^{(1)}$ or $s^{(1)}$ the diffusive eigenvalue.

We notice that the diffusive eigenvalue no longer exists beyond the
critical value $\beta_{\rm c}=\frac{1}{2}$. Since the matrix
(\ref{spec1aae}) has a total of $N$ eigenvalues, we expect
further nondiffusive eigenvalues in a number of $N-1$ for $\beta< \frac{1}{2}$
and $N$ for $\beta > \frac{1}{2}$, as confirmed in the following subsections.

\subsection{The eigenvalues $\mu^{(2)}$}

Beside the diffusive eigenvalue, we expect eigenvalues corresponding
to the chain-like
structure of the matrix (\ref{spec1aae}). To obtain these eigenvalues, we pose
$\tan \chi = 2i \beta \sin \theta $ so that Eqs. (\ref{spec1aal}),
(\ref{spec1aam}), and (\ref{spec1aan})
can be written respectively
\begin{eqnarray}
&& \sin(\theta N+\chi)=0 \; , \\ && \cos\left(\frac{\theta
N}{2}+\chi\right)=0 \; , \\ && \sin\left(\frac{\theta
N}{2}+\chi\right)=0 \; .
\end{eqnarray}
Therefore, we obtain
\begin{eqnarray}
i\left(n \pi - \theta N\right)&=&{\rm arctanh}(-2 \beta \sin \theta
) \; , \label{spec1aaq} \\
i\left(n\pi +\frac{\pi}{2} - \frac{\theta N}{2}\right)&=&{\rm
arctanh}(-2 \beta \sin \theta )\; , \label{spec1aar} \\
i\left(n \pi - \frac{\theta N}{2}\right)&=&{\rm arctanh}(-2 \beta
\sin \theta ) \; .\label{spec1abs}
\end{eqnarray}

We now expand ${\rm arctanh}(-2 \beta \sin \theta )$ around $\beta=0$:
\begin{eqnarray}
{\rm arctanh}(-2 \beta \sin \theta ) \stackrel{\beta \to 0}{=}
-2\beta \sin \theta - \frac{8}{3} \beta^3 \sin^3 \theta +
O(\beta^5) \; .\nonumber \\ \label{spec1aay}
\end{eqnarray}

If $\beta=0$, the solutions of Eqs. (\ref{spec1aaq}),
(\ref{spec1aar}), and (\ref{spec1abs})
are respectively given by
\begin{eqnarray}
\theta _0&=&\frac{n \pi}{N}, \ \ \text{where} \ \ n=1,2,\hdots,N-1 ,
\label{spec1aas} \\
\theta _0&=&\frac{(2n+1) \pi}{N}, \ \ \text{where} \ \
n=0,1,\hdots,\frac{N}{2}-1 , \label{spec1aat} \\
\theta _0&=&\frac{2n \pi}{N}, \ \ \text{where} \ \
n=1,2,\hdots,\frac{N}{2}-1 , \label{spec1aau}
\end{eqnarray}
Notice that $n=0$ is rejected in Eqs. (\ref{spec1aas}) and (\ref{spec1aau}).
It is due to the fact that $\theta =0$ does not correspond to an eigenvector
because it can be seen that $f_{l}=A+B\neq 0$ in Eq. (\ref{spec1aaf})
cannot be an eigenvector of Eq. (\ref{spec1aae}).

Using the expansion (\ref{spec1aay}) in Eqs. (\ref{spec1aaq}),
(\ref{spec1aar}) and (\ref{spec1abs}) with $\theta =\theta_0 + \delta
\theta $, we find
\begin{eqnarray}
\delta \theta \stackrel{\beta \to 0}{=} && - \frac{i 2 \beta}{M} \sin
\theta_0 - \frac{i 8 \beta^3}{3 M} \sin^3 \theta_0 \nonumber \\
&& + O\left(i \frac{\beta^5}{M}\right) + O\left(\frac{\beta^2}{M^2}\right) ,
\label{spec1aaz}
\end{eqnarray}
where $M=N$ for Eq. (\ref{spec1aaq}) and $M=\frac{N}{2}$ for Eqs.
(\ref{spec1aar}) and (\ref{spec1abs}).
The eigenvalue (\ref{spec1aag}) is now given by the expansion
\begin{eqnarray}
\mu &=& - 2i \beta \cos(\theta_0 + \delta \theta) \nonumber \\
&\stackrel{\delta \theta \to 0}{=}& - 2i \beta \cos \theta_0 + 2 i
\beta \sin \theta_0 \; \delta \theta \nonumber \\
&&+ i \beta \cos \theta_0 \; \delta \theta^2 + O(\delta \theta^3) .
\label{spec2aac}
\end{eqnarray}
Using Eq. (\ref{spec1aaz}) in (\ref{spec2aac}) gives
\begin{eqnarray}
\mu^{(2)} &\stackrel{\beta \to 0}{=}& - 2i \beta \cos \theta_0 +
O\left(i \frac{\beta^3}{M^2}\right)
\nonumber \\
&&+ \frac{4 \beta^2}{M} \sin^2 \theta_0 + \frac{16 \beta^4}{3 M}
\sin^4 \theta_0 +
O\left(\frac{\beta^6}{M}\right) \; . \nonumber \\ \label{spec2aad}
\end{eqnarray}
Consequently, we have
\begin{itemize}
\item for $N$ odd, using Eq. (\ref{spec2aad}) with (\ref{spec1aas}):
\begin{eqnarray}
\mu^{(2)} &\stackrel{\beta \to 0}{=}& - 2i \beta \cos \frac{n \pi}{N}
+ \frac{4 \beta^2}{N} \sin^2 \frac{n \pi}{N}
+ \frac{16 \beta^4}{3 N} \sin^4 \frac{n \pi}{N} , \nonumber \\ &&
\text{where} \ \
n=1,2,\hdots,N-1 \; ; \label{spec2aaf}
\end{eqnarray}
\item for $N=4I$ and $q$ even or for $N=4I+2$ and $q$ odd, using Eq.
(\ref{spec2aad}) with (\ref{spec1aat}):
\begin{eqnarray}
\mu^{(2)} &\stackrel{\beta \to 0}{=}& - 2i \beta \cos \frac{(2n+1)
\pi}{N} \nonumber \\ &&
+ \frac{8 \beta^2}{N} \sin^2 \frac{(2n+1) \pi}{N}
+ \frac{32 \beta^4}{3 N} \sin^4 \frac{(2n+1) \pi}{N}, \nonumber \\ &&
\text{where} \ \
n=0,1,\hdots,\frac{N}{2}-1 \; ; \label{spec2aai}
\end{eqnarray}
\item for $N=4I$ and $q$ odd or for $N=4I+2$ and $q$ even, using Eq.
(\ref{spec2aad})
with (\ref{spec1aau}):
\begin{eqnarray}
\mu^{(2)} &\stackrel{\beta \to 0}{=}& - 2i \beta \cos \frac{2n
\pi}{N} + \frac{8 \beta^2}{N} \sin^2 \frac{2n \pi}{N}
+ \frac{32 \beta^4}{3 N} \sin^4 \frac{2n \pi}{N}, \nonumber \\ &&
\text{where} \ \
n=1,2,\hdots,\frac{N}{2}-1 \; .\label{spec2aal}
\end{eqnarray}
\end{itemize}

\subsection{The eigenvalues $\mu^{(3)}$}

The solutions of Eq. (\ref{spec1abm}) are simply given by
\begin{eqnarray}
\theta =\frac{2 n \pi}{N}, \ \ \text{where} \ \
n=1,2,\hdots,\frac{N}{2}-1\; . \label{spec1aao}
\end{eqnarray}
We reject $n=0$ and $n=\frac{N}{2}$ because the corresponding
eigenvector does not exist in these cases. Similarly, the solutions
of Eq. (\ref{spec1abn}) are given by
\begin{eqnarray}
\theta =\frac{(2n+1) \pi}{N}, \ \ \text{where} \ \
n=0,1,\hdots,\frac{N}{2}-1\; . \label{spec1aap}
\end{eqnarray}

Consequently, we have the further eigenvalues:\\
\begin{itemize}
\item for $N=4I$ and $q$ even or for $N=4I+2$ and $q$ odd, using Eq.
(\ref{spec1aag}) with (\ref{spec1aao}):
\begin{eqnarray}
\mu^{(3)} &=& - 2i \beta \cos\left(\frac{2n \pi}{N}\right) \; , \ \
\text{where} \ \ n=1,2,\hdots,\frac{N}{2}-1 \; ;
\nonumber \\ \label{spec2aah}
\end{eqnarray}
\item for $N=4I$ and $q$ odd or for $N=4I+2$ and $q$ even, using Eq.
(\ref{spec1aag}) with (\ref{spec1aap}):
\begin{eqnarray}
\mu^{(3)} &=& - 2i \beta \cos\left(\frac{(2n+1) \pi}{N}\right), \nonumber \\
&& \text{where} \ \ n=0,1,\hdots,\frac{N}{2}-1 \; .
\label{spec2aak}
\end{eqnarray}
\end{itemize}

\subsection{The eigenvalues $\mu^{(4)}$}

An important observation is that,
for $\beta < \frac{1}{2}$, the expansion (\ref{spec1aay}) which
implies $2 \beta \sin \theta < 1$
is satisfied everywhere, i.e., for all the values of $\theta$ and
therefore for all the eigenvalues.
However, for $\beta > \frac{1}{2}$, the Taylor expansion around
$\beta=0$ in Eq. (\ref{spec1aay}) is only valid if $2 \beta \sin
\theta < 1$.
Therefore, a transition zone exists around $\sin \theta \simeq 1/(2 \beta)$.
According to Eq. (\ref{spec1aag}), this transition corresponds to the
critical value of the eigenvalue
given by
\begin{eqnarray}
\mu_{\rm c} &=& \pm i \sqrt{(2 \beta)^2-1} \; . \label{spec2abc}
\end{eqnarray}
Therefore, for $\beta > \frac{1}{2}$, the expansion (\ref{spec1aay})
around $\beta=0$ is only valid if $\vert \mu \vert > \vert \mu_{\rm
c}\vert$.
For $\vert \mu \vert < \vert \mu_{\rm c}\vert$, we should instead
consider the asymptotic expansion of ${\rm arctanh}(-2 \beta \sin
\theta )$ around $\beta=\infty$:
\begin{eqnarray}
{\rm arctanh}(-2 \beta \sin \theta ) &\stackrel{\beta \to \infty}{=}&
i \frac{\pi}{2} -\frac{1}{2 \beta \sin \theta } -\frac{1}{24 \beta^3
\sin^3 \theta } \nonumber \\
&& -\frac{1}{160 \beta^5 \sin^5 \theta } +
O\left(\frac{1}{\beta^7}\right) \; , \nonumber \\
\label{spec2aaa}
\end{eqnarray}
which leads to another family of eigenvalues existing for $\beta >
\frac{1}{2}$.

If $\beta\to\infty$, the solutions of Eqs. (\ref{spec1aaq}),
(\ref{spec1aar}) and (\ref{spec1abs})
are respectively given by
\begin{eqnarray}
\theta _0&=&\frac{(n+\frac{1}{2}) \pi}{N}, \ \ \text{where} \ \
n=0,1,\hdots,N-1 \, , \label{spec1aav} \\
\theta _0&=&\frac{2n \pi}{N}, \ \ \text{where} \ \
n=1,2,\hdots,\frac{N}{2}-1 \, , \label{spec1aaw} \\
\theta _0&=&\frac{(2n+1) \pi}{N}, \ \ \text{where} \ \
n=0,1,\hdots,\frac{N}{2}-1 . \label{spec1aax}
\end{eqnarray}
Because of the condition $\vert \mu \vert < \vert \mu_{\rm c}\vert$
with the critical values (\ref{spec2abc}),
we should only consider the angles in the interval $\theta_{0,{\rm
c}}< \theta_0 < \pi - \theta_{0,{\rm c}}$ with
\begin{eqnarray}
\theta_{0,{\rm c}} &=& \arcsin \frac{1}{2\beta} \; ,
\label{theta0c}
\end{eqnarray}
so that the integer $n$ in Eqs. (\ref{spec1aav}), (\ref{spec1aaw}),
and (\ref{spec1aax})
is restricted to take the intermediate values $n_{\rm min} < n <
n_{\rm max}$ which do
not reach the extreme values.

Using the expansion (\ref{spec2aaa}) in Eqs. (\ref{spec1aaq}),
(\ref{spec1aar}), and (\ref{spec1abs}) with $\theta =\theta_0 +
\delta \theta $, we find
\begin{eqnarray}
\delta \theta &\stackrel{\beta\to\infty}{=}& - \frac{i}{2 \beta M
\sin \theta_0}
- \frac{i}{24 \beta^3 M \sin^3 \theta_0} - \frac{i}{160 \beta^5 M
\sin^5 \theta_0}\nonumber \\
&&+ O\left(\frac{i}{\beta^7 M}\right) + O\left(\frac{1}{\beta^2
M^2}\right) \; , \label{spec2aab}
\end{eqnarray}
where $M=N$ for Eq. (\ref{spec1aaq}) and $M=\frac{N}{2}$ for Eqs.
(\ref{spec1aar}) and (\ref{spec1abs}).

Using Eq. (\ref{spec2aab}) in (\ref{spec2aac}) gives
\begin{eqnarray}
\mu^{(4)} &\stackrel{\beta \to \infty}{=}& - 2i \beta \cos \theta_0 +
\frac{1}{M} \left(1+
\frac{1}{12 \beta^2 \sin^2 \theta_0 } \right. \nonumber \\ &&\left. +
\frac{1}{80 \beta^4 \sin^4 \theta_0}\right)
+ O\left(\frac{1}{\beta^6 M}\right) + O\left(\frac{i}{\beta
M^2}\right) \; . \nonumber\\ \label{spec2aae}
\end{eqnarray}
Consequently, we have
\begin{itemize}
\item for $N$ odd, using Eq. (\ref{spec2aae}) with (\ref{spec1aav}):
\begin{eqnarray}
\mu^{(4)} &\stackrel{\beta \to \infty}{=}& - 2i \beta \cos
\frac{(n+\frac{1}{2}) \pi}{N}
+\frac{1}{N} \left(1+\frac{1}{12 \beta^2 \sin^2 \frac{(n+\frac{1}{2})
\pi}{N}} \right. \nonumber \\
&&\left.
+\frac{1}{80\beta^4\sin^4 \frac{(n+\frac{1}{2}) \pi}{N}}\right)
, \nonumber \\ && \text{where} \ \ n=0,1,\hdots,N-1 \; ;\label{spec2aag}
\end{eqnarray}
\item for $N=4I$ and $q$ even or for $N=4I+2$ and $q$ odd, using Eq.
(\ref{spec2aae}) with (\ref{spec1aaw}):
\begin{eqnarray}
\mu^{(4)} &\stackrel{\beta \to \infty}{=}& - 2i \beta \cos \frac{2 n
\pi}{N} \nonumber \\
&&+ \frac{2}{N} \left(1+\frac{1}{12 \beta^2 \sin^2 \frac{2 n \pi}{N}}
+\frac{1}{80 \beta^4 \sin^4 \frac{2 n \pi}{N}}\right)
, \nonumber \\ && \text{where} \ \ n=1,2,\hdots,\frac{N}{2}-1 \;
;\label{spec2aaj}
\end{eqnarray}
\item for $N=4I$ and $q$ odd or for $N=4I+2$ and $q$ even, using Eq.
(\ref{spec2aae}) with (\ref{spec1aax}):
\begin{eqnarray}
\mu^{(4)} &\stackrel{\beta \to \infty}{=}& - 2i \beta \cos \frac{(2
n+1) \pi}{N} \nonumber \\
&&+ \frac{2}{N} \left(1+\frac{1}{12 \beta^2 \sin^2 \frac{(2 n+1)
\pi}{N}} \right.
\nonumber \\
&&\left. +\frac{1}{80 \beta^4 \sin^4 \frac{(2 n+1) \pi}{N}}\right)
, \nonumber \\ && \text{where} \ \ n=0,1,\hdots,\frac{N}{2}-1 \;
;\label{spec2aam}
\end{eqnarray}
\end{itemize}
with the aforementioned restriction on the values of the integer $n$.

\subsection{The eigenvalues $\mu^{(5)}$}

For $N=4I$ and $q$ even or $N=4I+2$ and $q$ odd,
two special eigenvalues exist in the limit $\beta\to\infty$ around
$\theta_0=0$ and $\theta_0=\pi$.
They can be obtained by taking $\theta=\theta_0+\delta \theta$ and
directly solving
Eq. (\ref{spec1aam}) to get in both cases
\begin{eqnarray}
\delta \theta^2 \stackrel{\beta\to\infty}{=} \frac{1}{i \beta N} +
O\left(\frac{1}{\beta^2}\right) \; . \label{spec1abo}
\end{eqnarray}
Inserting in Eq. (\ref{spec2aac}), we obtain
\begin{eqnarray}
\mu^{(5)} &\stackrel{\beta \to \infty}{=}& \mp 2i \beta + \frac{1}{N}
+ O\left(\frac{1}{\beta}\right) \; .
\label{spec2abh}
\end{eqnarray}


\subsection{Description of the spectrum}
\label{spectradisuss}

Rewriting the eigenvalues (\ref{spec1aag}) of the Redfield
superoperator with their explicit dependence
in terms of Eq. (\ref{spec1bab}), we get
\begin{eqnarray}
\mu_{\nu} = \mu_{q\theta} = -2 i \beta \cos \theta =
-2 i \frac{A}{Q \lambda^2} \left(\sin \frac{q}{2}\right) \cos \theta
\; ,\label{spec3abd}
\end{eqnarray}
where $\nu$ can take $N^2$ different values because $q$ and $\theta$
take $N$ values each.
Remembering that according to Eq. (\ref{spec1baa})
\begin{eqnarray}
s_{\nu} = s_{q\theta} = 2 Q \lambda^2 (\mu_{q\theta}- 1), \label{spec3acd}
\end{eqnarray}
we conclude that we have found in this section all the eigenvalues of
the Redfield superoperator.
We list them in Tables \ref{table1} and \ref{table2}
according to the parameter regime in which they hold.

We now discuss the main features of the spectrum when the different
physical parameters are varied.
This discussion is based on our analytical results for the eigenvalues
and on the comparison between these
results and the eigenvalues obtained by numerical diagonalization of
the Redfield superoperator.
Since the eigenvalues $\mu_{q\theta}$ are related to the Redfield
superoperator eigenvalues $s_{q\theta}$ by Eq. (\ref{spec3acd}), we notice that
all the eigenvalues of the complete spectrum always satisfy $0\leq
{\rm Re} \; \mu_{q\theta} \leq 1$ or, equivalently, $-2 Q \lambda^2
\leq {\rm Re} \; s_{q\theta} \leq 0$.
The imaginary part of $s_{q\theta}$ is simply proportional by a
factor $2 Q \lambda^2$
to the imaginary part of $\mu_{q\theta}$.

We start by studying the $N$ eigenvalues $\mu_{q\theta}$ obtained by
fixing the wavenumber
$q$ (even or odd) and varying $\theta$.
For given physical parameters ($A$, $\lambda$, $Q$, $N$), fixing $q$
is equivalent to fixing $\beta$.

\begin{figure}[h]
\centering
\begin{tabular}{c@{\hspace{0.5cm}}c}
\vspace*{1cm}
\rotatebox{0}{\scalebox{0.35}{\includegraphics{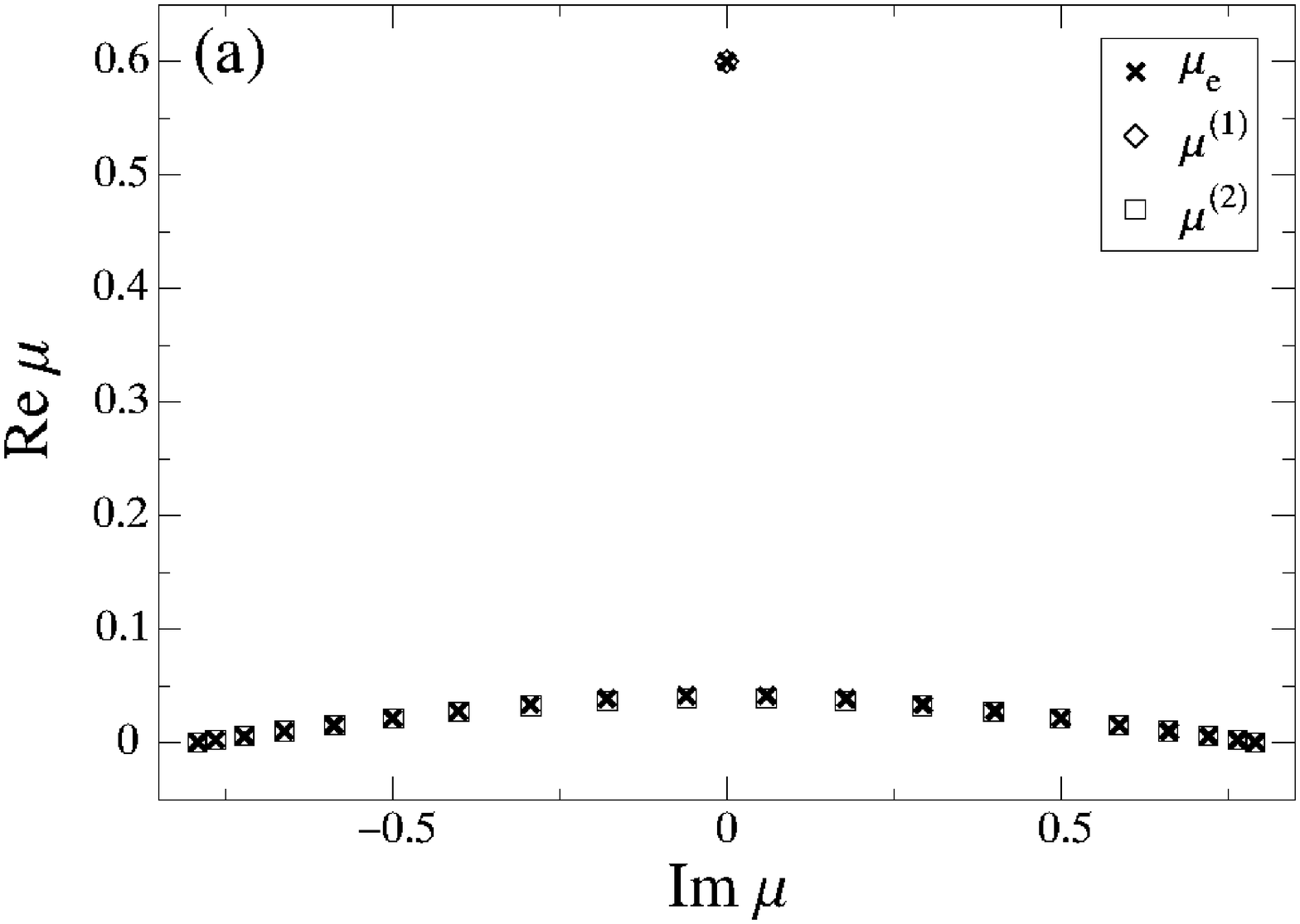}}} \\
\vspace*{1cm}
\rotatebox{0}{\scalebox{0.35}{\includegraphics{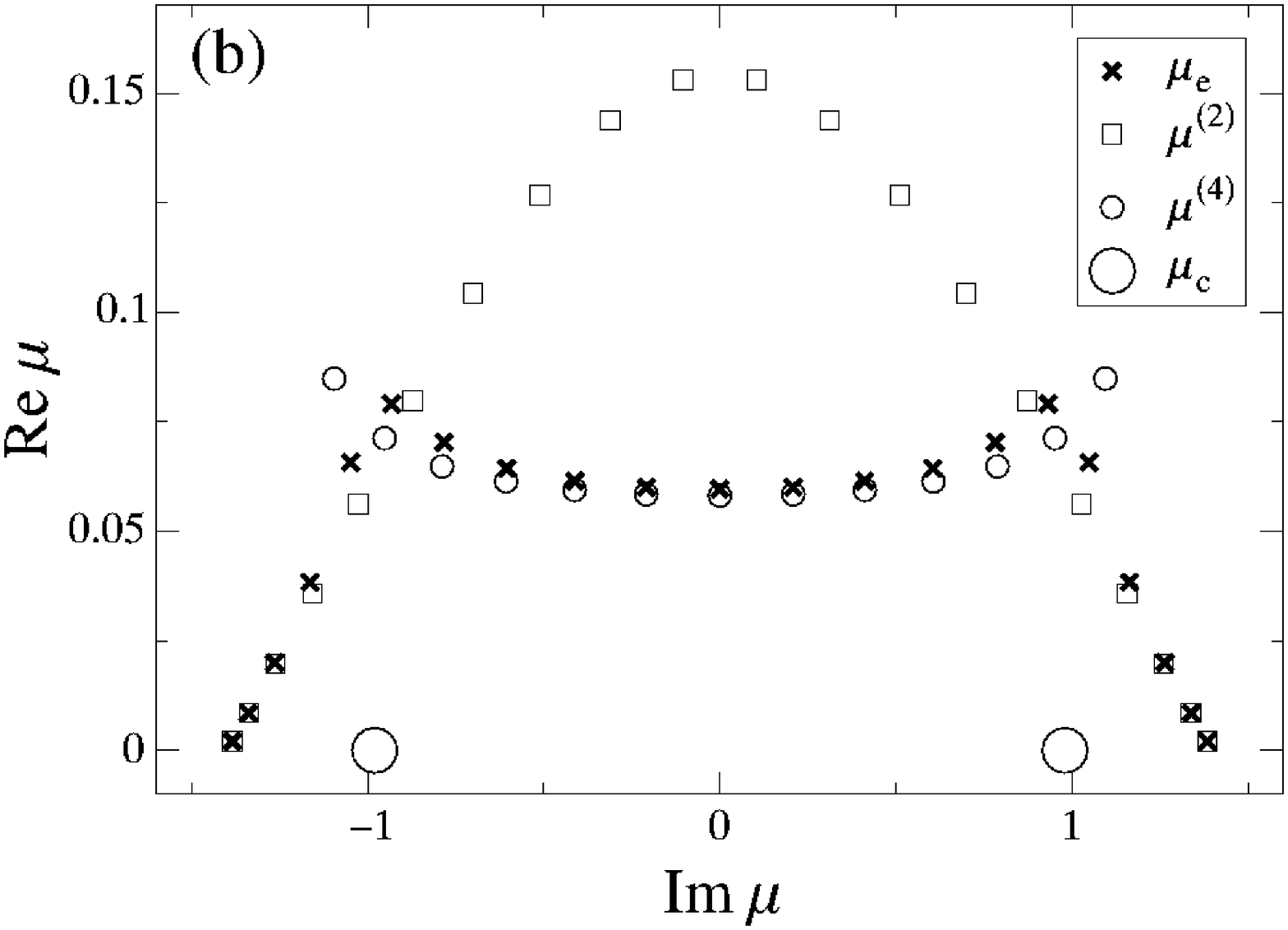}}} \\
\vspace*{1cm}
\rotatebox{0}{\scalebox{0.35}{\includegraphics{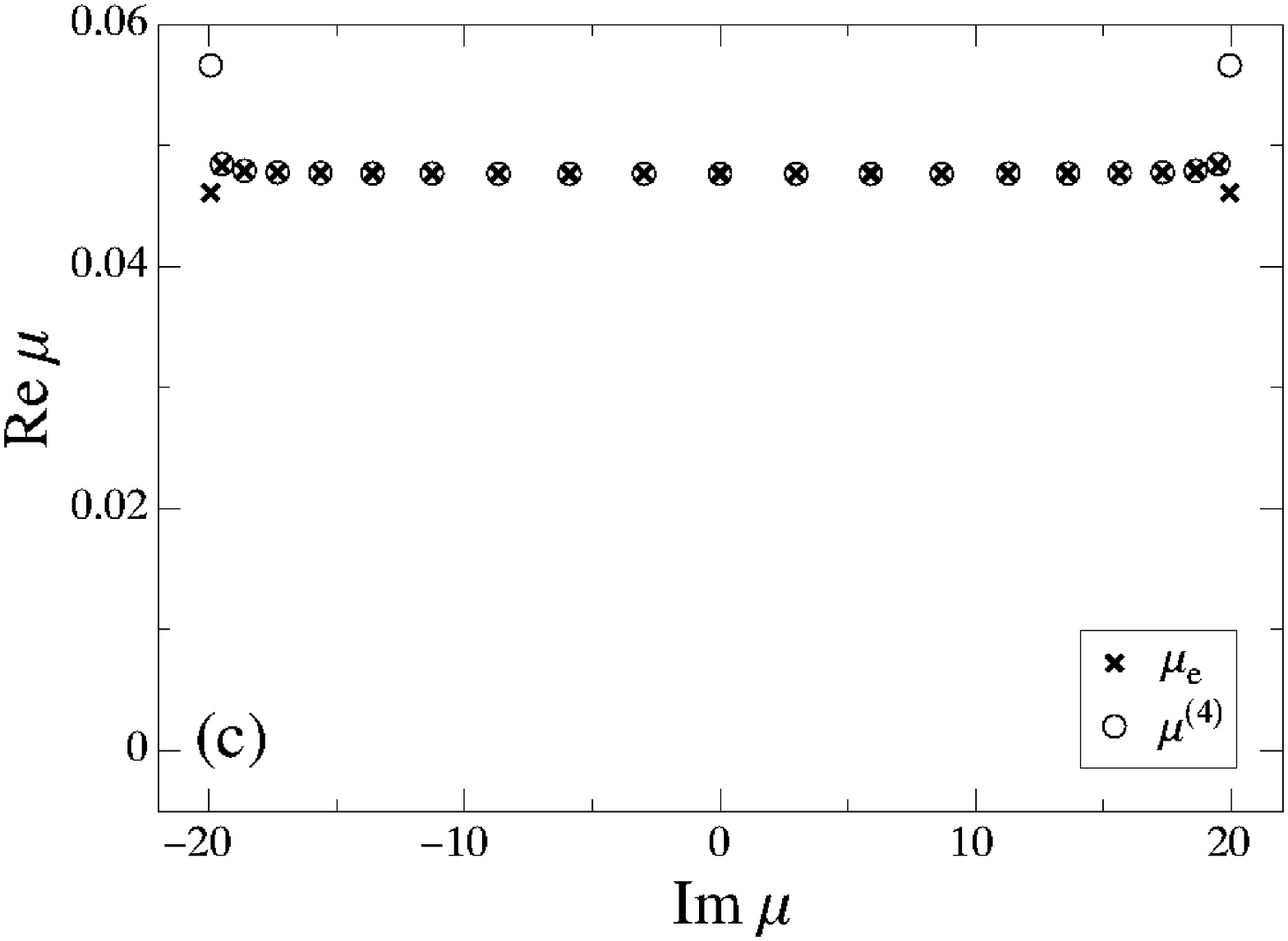}}} \\
\end{tabular}
\caption{Eigenvalue spectrum for $N=21$ and given $q$: (a) $\beta=0.4$,
(b) $\beta=0.7$, and (c) $\beta=10$. $\mu_{\rm e}$ denotes the exact
eigenvalues obtained
by numerical diagonalization and $\mu^{(i)}$ the eigenvalues of the
different families
given in Tables \ref{table1} and \ref{table2}}. \label{spectre2D.Nodd}
\end{figure}

\begin{figure}[h]
\centering
\begin{tabular}{c@{\hspace{0.5cm}}c}
\vspace*{1cm}
\rotatebox{0}{\scalebox{0.35}{\includegraphics{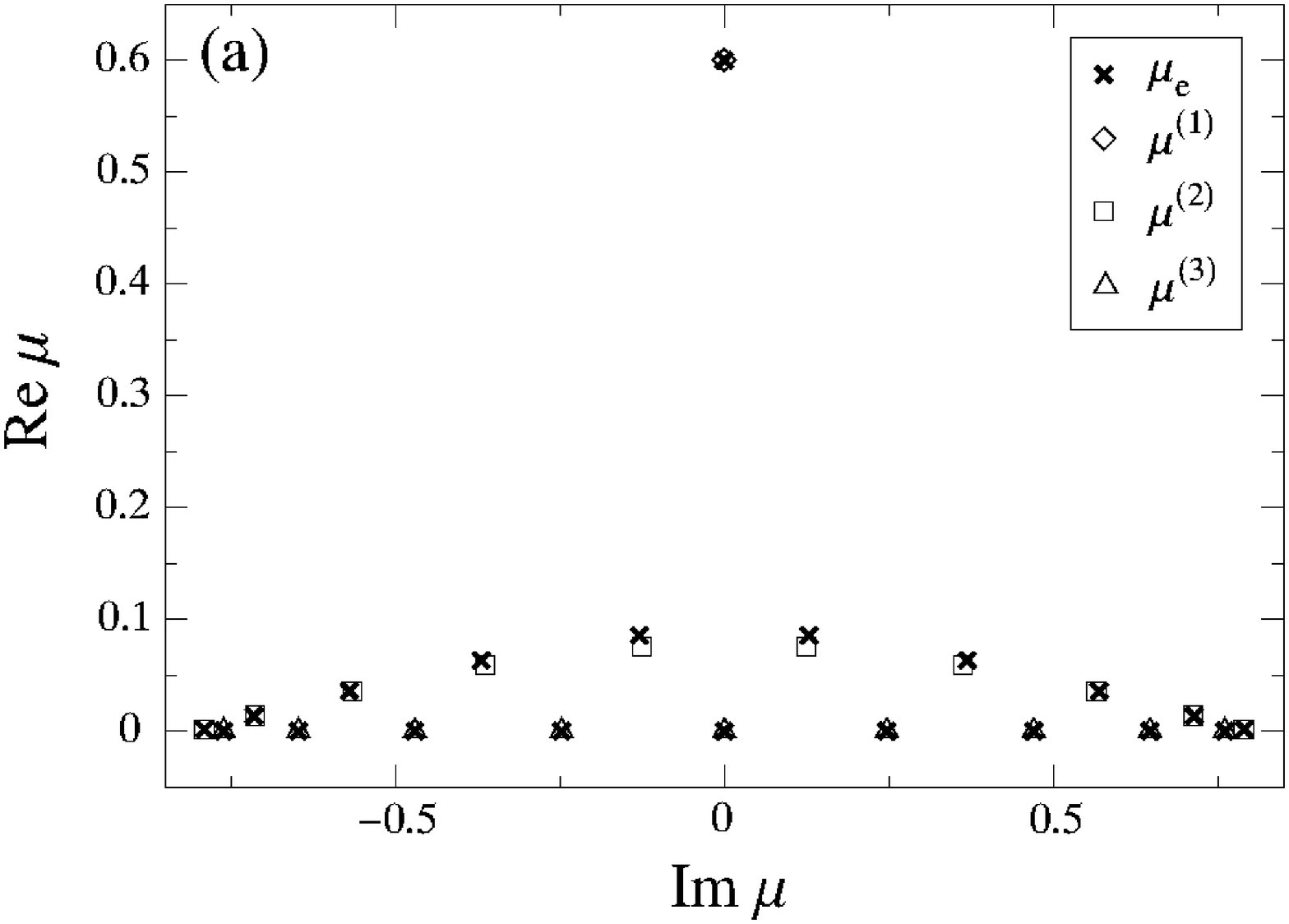}}} \\
\vspace*{1cm}
\rotatebox{0}{\scalebox{0.35}{\includegraphics{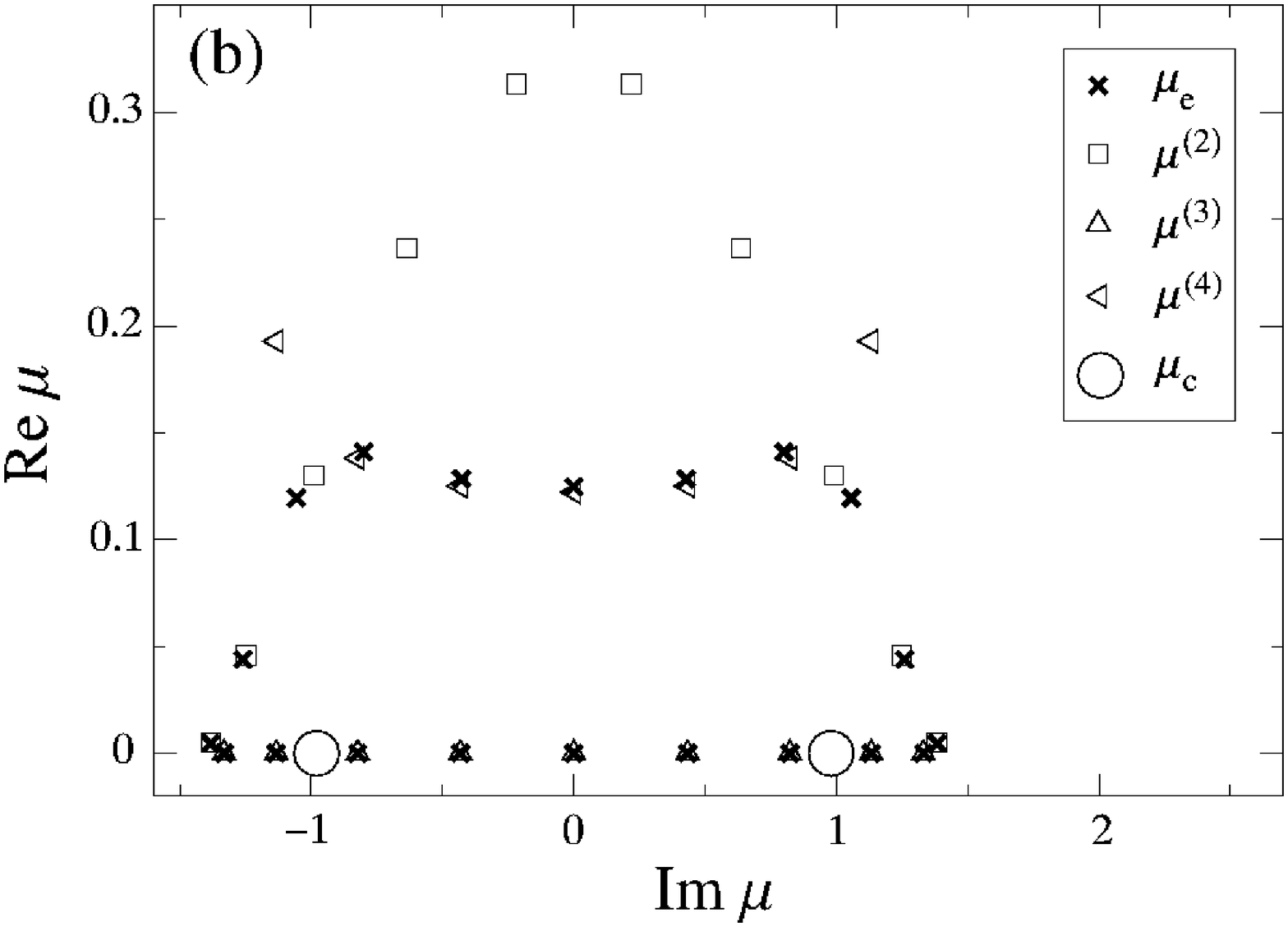}}} \\
\vspace*{1cm}
\rotatebox{0}{\scalebox{0.35}{\includegraphics{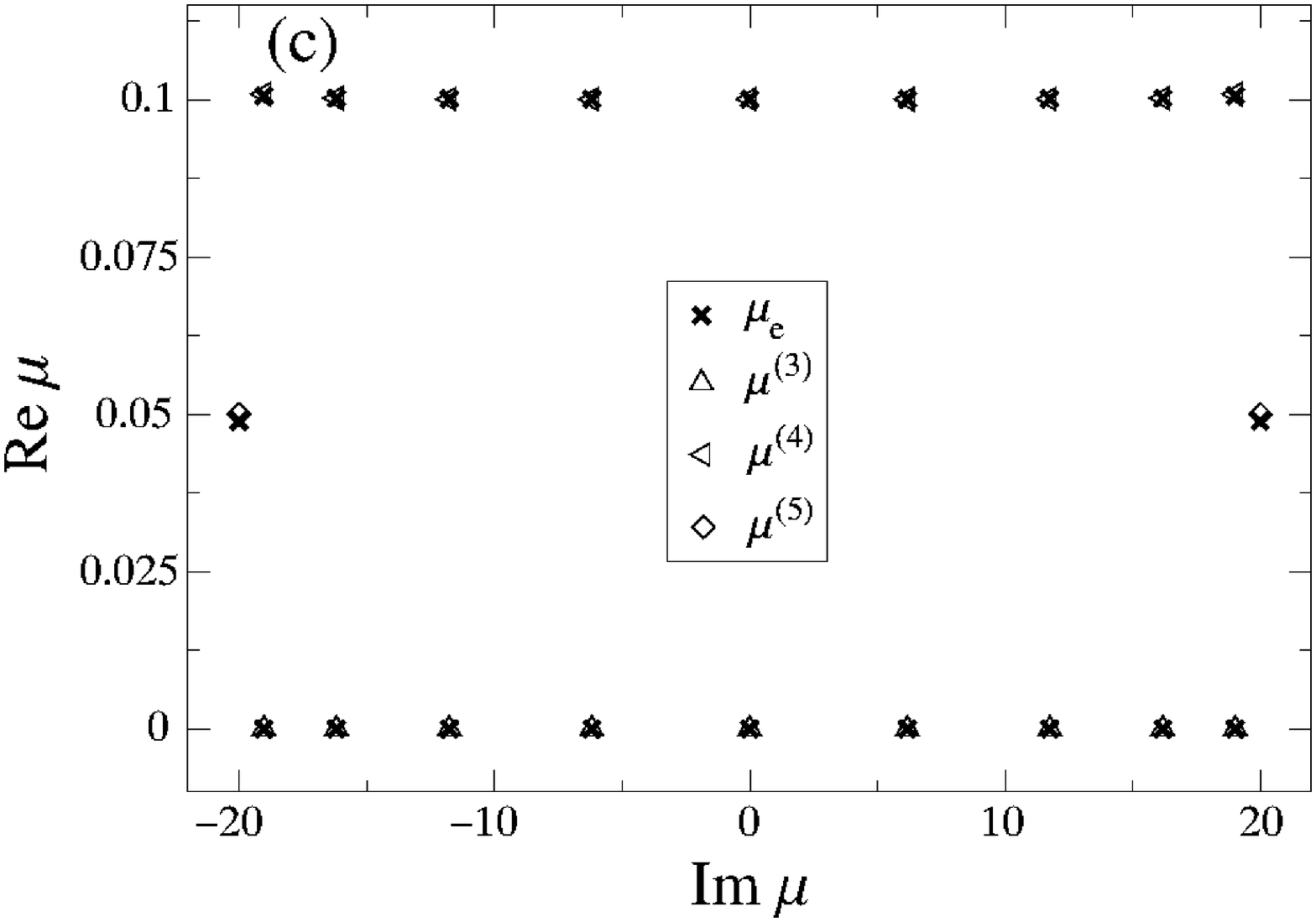}}} \\
\end{tabular}
\caption{Eigenvalue spectrum for $N=20$ and $q$ even: (a) $\beta=0.4$,
(b) $\beta=0.7$, and (c) $\beta=10$. $\mu_{\rm e}$ denotes the exact
eigenvalues obtained
by numerical diagonalization and $\mu^{(i)}$ the eigenvalues of the
different families
given in Tables \ref{table1} and \ref{table2}}. \label{spectre2D.Neven.qeven}
\end{figure}

\begin{figure}[h]
\centering
\begin{tabular}{c@{\hspace{0.5cm}}c}
\vspace*{1cm}
\rotatebox{0}{\scalebox{0.35}{\includegraphics{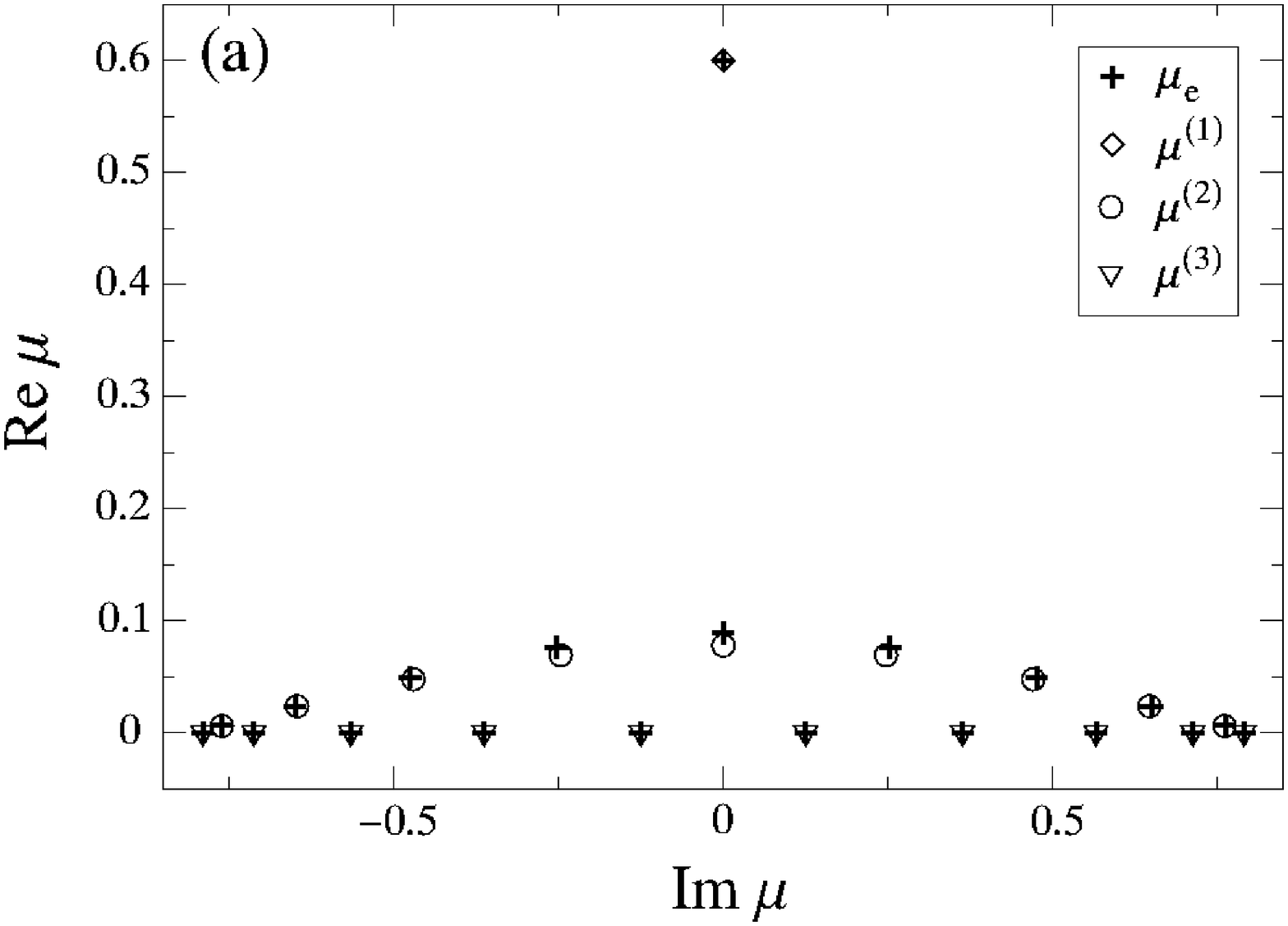}}} \\
\vspace*{1cm}
\rotatebox{0}{\scalebox{0.35}{\includegraphics{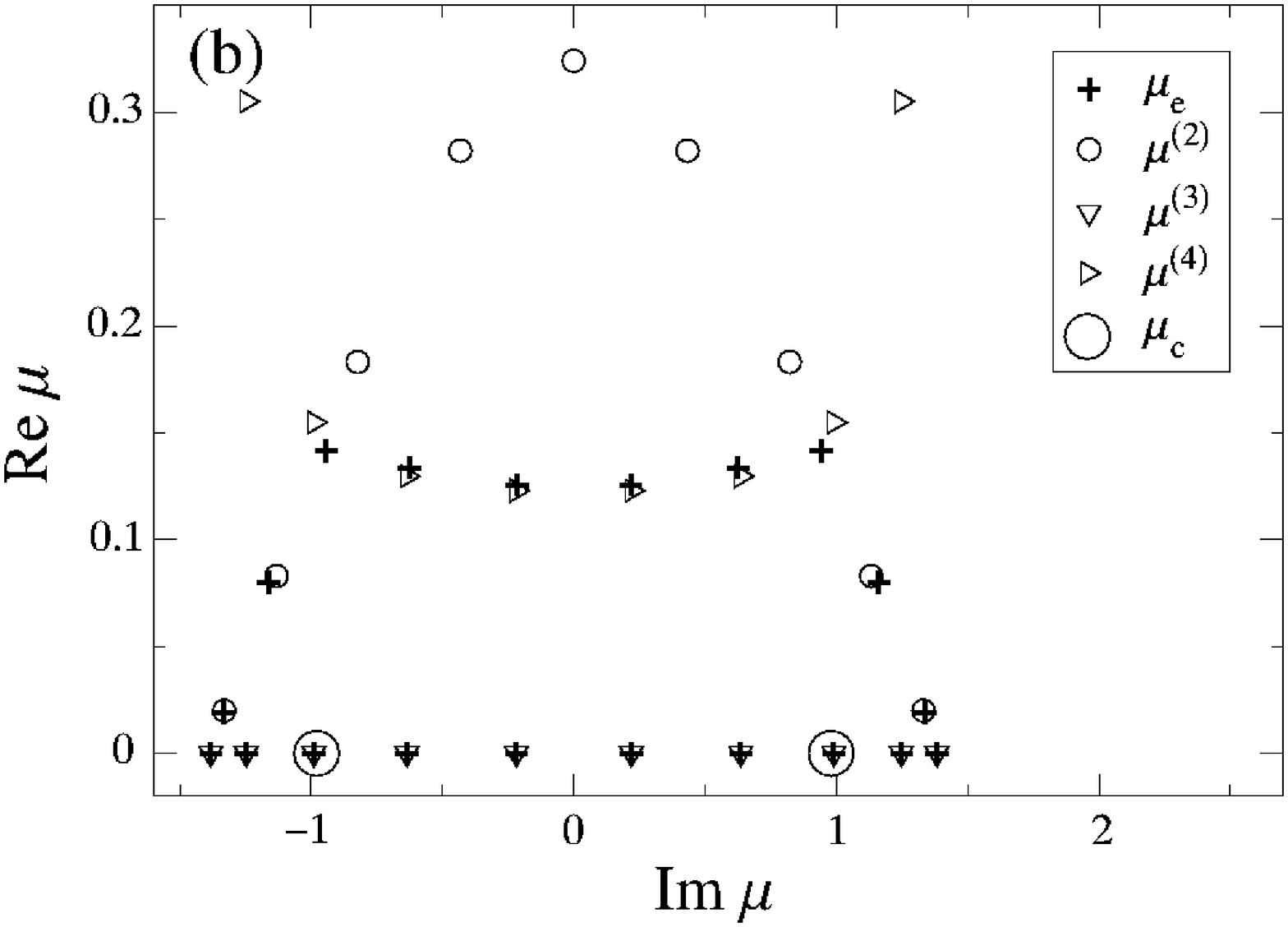}}} \\
\vspace*{1cm}
\rotatebox{0}{\scalebox{0.35}{\includegraphics{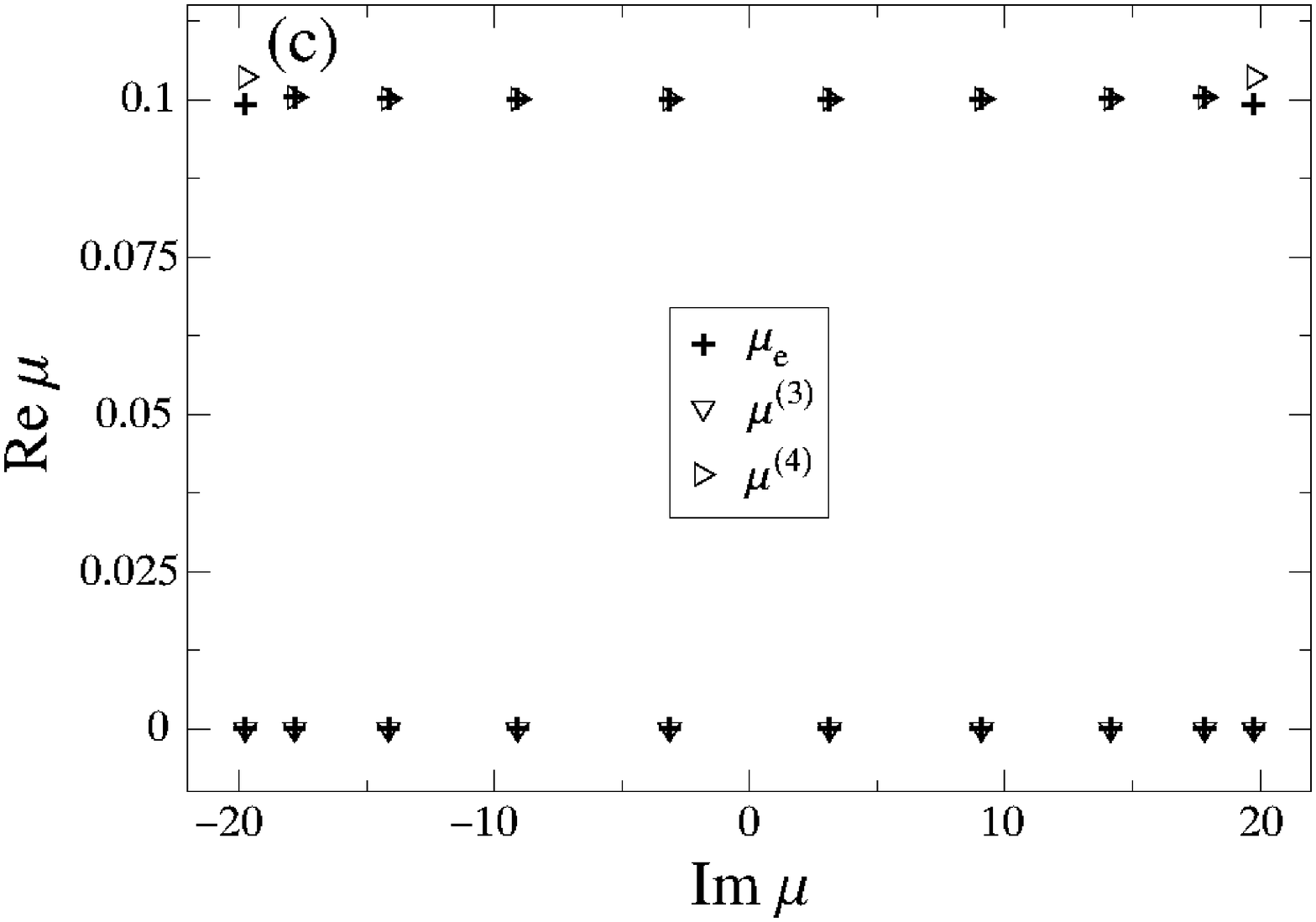}}} \\
\end{tabular}
\caption{Eigenvalue spectrum for $N=20$ and $q$ odd: (a) $\beta=0.4$,
(b) $\beta=0.7$, and (c) $\beta=10$. $\mu_{\rm e}$ denotes the exact
eigenvalues obtained
by numerical diagonalization and $\mu^{(i)}$ the eigenvalues of the
different families
given in Tables \ref{table1} and \ref{table2}}. \label{spectre2D.Neven.qodd}
\end{figure}

For $\beta < \frac{1}{2}$, the analytical expressions of the
eigenvalues which concern us are
summarized in Table \ref{table1}.
Two families of eigenvalues ($\mu^{(1)}$ and $\mu^{(2)}$) enter in
the discussion for $N$ odd, and three families ($\mu^{(1)}$,
$\mu^{(2)}$, and $\mu^{(3)}$), for $N$ even.
The numerical eigenvalues are plotted in Figs. \ref{spectre2D.Nodd}(a),
\ref{spectre2D.Neven.qeven}(a), and \ref{spectre2D.Neven.qodd}(a)
and are in very good agreement with the analytical results.
The sole diffusive eigenvalue $\mu^{(1)}$ has a real part and no
imaginary part.
The $N-1$ other eigenvalues, either belongs to the $\mu^{(2)}$ family
for $N$ odd or to the $\mu^{(2)}$
and $\mu^{(3)}$ families for $N$ even.
The eigenvalues $\mu^{(2)}$ and $\mu^{(3)}$ have an imaginary part
which extends from $-2\beta$ to
$2\beta$ and they generate oscillations in the dynamics.
The real part of the $\mu^{(2)}$ eigenvalues is small and tends to
zero in the large $N$ limit.
The real part of the $\mu^{(3)}$ eigenvalues is always zero.

For $\beta > \frac{1}{2}$, the diffusive eigenvalue $\mu^{(1)}$ has
disappeared
after merging with the other eigenvalues and
the situation is slightly more complicated.
The situation for a moderate value of $\beta > \frac{1}{2}$ is
depicted in Figs. \ref{spectre2D.Nodd}(b),
\ref{spectre2D.Neven.qeven}(b), and \ref{spectre2D.Neven.qodd}(b)
while the analytical expressions of the eigenvalues are given
in Table \ref{table2}.
Since $\mu^{(1)}$ no longer exists, we have
the two families of eigenvalues $\mu^{(2)}$ and $\mu^{(4)}$ if $N$ is odd, and
the three families $\mu^{(2)}$, $\mu^{(3)}$ and $\mu^{(4)}$ if $N$ is even.
Two regions of the spectrum have to be distinguished.
The eigenvalues $\mu^{(2)}$ exist in the region where $\vert \mu
\vert > \vert \mu_c\vert$ while the eigenvalues $\mu^{(4)}$ exist in
the region where $\vert \mu \vert <\vert \mu_c\vert$.
We observe that the extra family of eigenvalues $\mu^{(4)}$ has
appeared because
of the collision with the diffusive eigenvalue $\mu^{(1)}$.
We can see in Figs. \ref{spectre2D.Nodd}(b),
\ref{spectre2D.Neven.qeven}(b), and \ref{spectre2D.Neven.qodd}(b)
that the analytical results of Table \ref{table2}
reproduce very well the eigenvalues obtained by numerical
diagonalization in the two regions.
Here again, the number of eigenvalues is equal to $N$ for a given
wavenumber $q$,
the imaginary part of the eigenvalues extends from $-2 \beta$ to $2 \beta$, and
the real parts of all eigenvalues tends to zero in the large $N$ limit.

A special situation occurs when $\beta> \frac{1}{2}$ is increased to
large values.
This situation is depicted in Figs. \ref{spectre2D.Nodd}(c),
\ref{spectre2D.Neven.qeven}(c), and \ref{spectre2D.Neven.qodd}(c).
The situation is similar to the previous one but the region $\vert
\mu \vert > \vert\mu_c\vert$ has disappeared so that the family of
eigenvalues $\mu^{(2)}$ corresponding to the expansion $\beta\to 0$
no longer exists.
For $N$ odd and for $N$ even with $q$ odd, these eigenvalues are
replaced by the eigenvalues $\mu^{(4)}$ eigenvalues. For $N$ even
and $q$ even, we find the two eigenvalues $\mu^{(5)}$ beside the
family of eigenvalues $\mu^{(4)}$. The agreement between the
analytical and numerical results is very good here also.
As before, the imaginary part of all these eigenvalues extends from
$-2 i \beta$ to $2 i \beta$ and
their real parts tends to zero in the large $N$ limit.

\begin{figure}[h]
\centering
\begin{tabular}{c@{\hspace{0.5cm}}c}
\rotatebox{0}{\scalebox{0.8}{\includegraphics{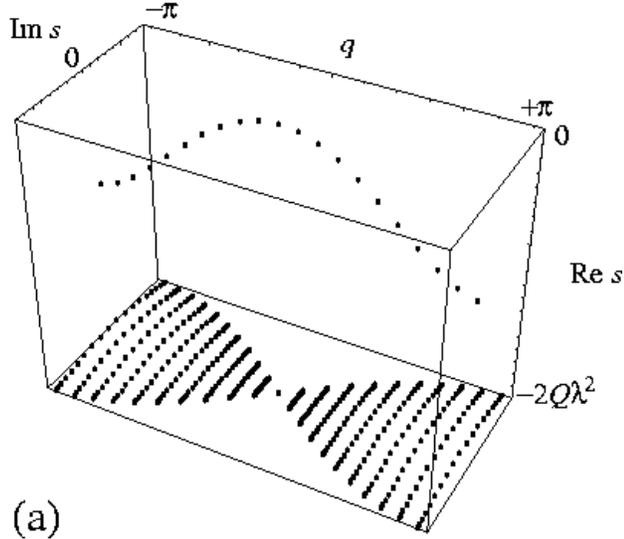}}} \\
\rotatebox{0}{\scalebox{0.8}{\includegraphics{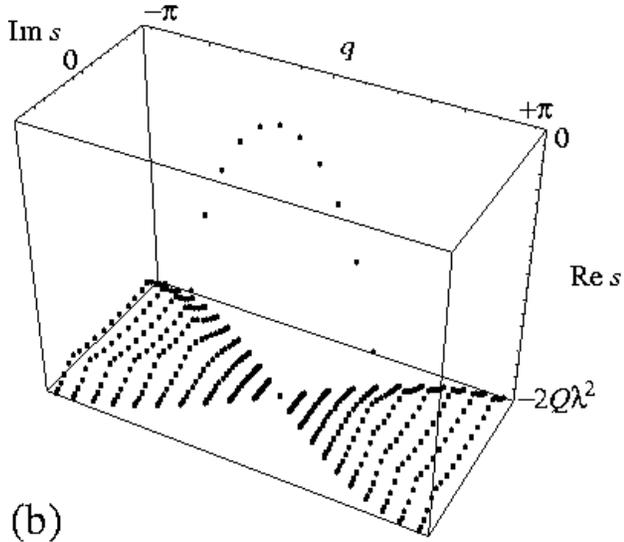}}} \\
\rotatebox{0}{\scalebox{0.8}{\includegraphics{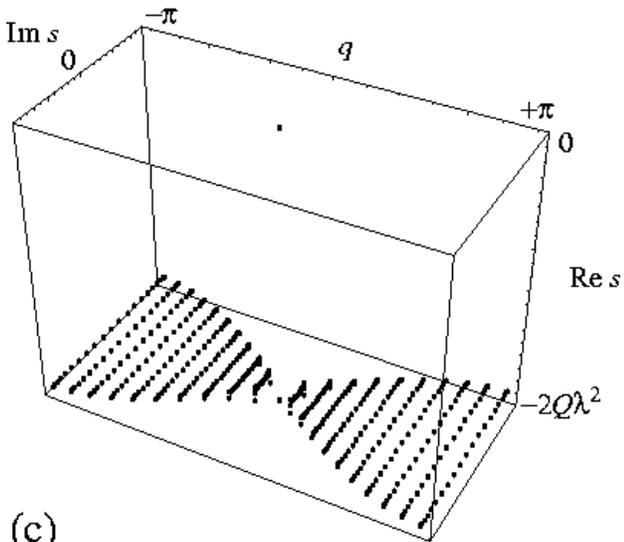}}} \\
\end{tabular}
\caption{Complete spectrum of the eigenvalues $s_{q\theta}$ 
versus the wavenumber $q$ 
for the Redfield superoperator (\ref{isolaam})-(\ref{blochaaa})
The eigenvalues are given by Eq. (\ref{spec3acd})
in terms of the eigenvalues $\mu_{q\theta}$ depicted
in Figs. \ref{spectre2D.Nodd}-\ref{spectre2D.Neven.qodd}  
at given values of the wavenumber.
The size of the chain is here $N=21$. The parameter values are: 
(a) $\frac{A}{Q \lambda^2}=0.4$;
(b) $\frac{A}{Q \lambda^2}=0.7$; (c) $\frac{A}{Q
\lambda^2}=10$.} \label{spectre3D}
\end{figure}

A global view of the complete spectrum of the $N^2$ eigenvalues of
the Redfield superoperator
is depicted in Fig. \ref{spectre3D} by varying the wavenumber $q$ in
a third dimension.
Here, we only consider for simplicity the case where $N$ is odd.
The relation between the wavenumber $q$ and the parameter $\beta$ is
given by Eq. (\ref{spec1bab}).
The wavenumber $q$ varies in the first Brillouin zone or,
equivalently, in the interval $0\leq q < 2\pi$.
We see in Fig. \ref{spectre3D}(a) that the diffusive eigenvalues
$\mu^{(1)}$ exists for
all the values of the wavenumber in the case $\frac{A}{Q \lambda^2}<
\frac{1}{2}$
which implies $\beta < \frac{1}{2}$.
However, if $\frac{A}{Q \lambda^2}> \frac{1}{2}$, the diffusive
eigenvalue $\mu^{(1)}$ disappears as expected for some values of the
wavenumber corresponding to $\beta > \frac{1}{2}$.
This situation is observed in Fig. \ref{spectre3D}(b).

For very large values of $\frac{A}{Q \lambda^2}> \frac{1}{2}$, the
diffusive branch of the
spectrum is reduced to the sole eigenvalue at $q^{(1)}=0$, as seen in
Fig. \ref{spectre3D}(c).
In this case, diffusion has disappeared from the spectrum which only contains
eigenvalues associated with damped oscillatory behavior.
The diffusive branch can be supposed to have disappeared when its
last nonzero eigenvalue disappears in Eq. (\ref{spec3aae}).
Therefore the diffusive branch disappears when the value of $\beta$
for the first nonzero eigenvalue
corresponding to $q=\frac{2\pi}{N}$ is larger than the critical value
$\beta_{\rm c}=\frac{1}{2}$.
This happens when the coupling parameter exceeds the critical value given by
\begin{eqnarray}
\lambda_{\rm c} = \sqrt{\frac{2 A}{Q} \sin \frac{\pi}{N}}
\stackrel{N > 5}{\simeq}
\sqrt{\frac{2 A \pi}{Q N}} \; . \label{spec3aah}
\end{eqnarray}
This disappearance of the diffusion branch can be observed in Fig.
\ref{spectre3D}(c).
We notice that the diffusive branch always exists in the
infinite-system limit ($N \to \infty$) in which case $\lambda_{\rm
c}$ can be arbitrarily small.

We have described in this section the complete spectrum of the
Redfield superoperator for a finite chain. 
We now briefly indicate which are the dynamical implications of these 
results.


\subsection{From the spectrum to the dynamics}
\label{dynamicconsequence}

The linear decomposition (\ref{isolaan}) of the density matrix shows
that the modes which control the long-time dynamics correspond to the eigenvalues 
having the smallest absolute value of their real part. The eigenvalues with larger absolute 
value of their real part correspond to faster modes which control the relaxation on shorter 
time scales.  For systems of finite size $N$, the long-time relaxation can be of two kinds.
In the case where $\lambda<\lambda_{\rm c}$, the long-time relaxation is 
controlled by nondiffusive modes and consists in complicated oscillations of different periods 
damped at rates ${\rm Re}\, s \simeq -2 Q \lambda^2$. 
This is due to the fact that, in this case, all the eigenvalues of the spectrum have a similar 
real part and different imaginary parts. 
This also indicates that the modes corresponding to the relaxation of the coherences (modes 
with complex eigenvalues) and of the populations (modes with real eigenvalues) decay on 
similar time scales.
In the other case where $\lambda>\lambda_{\rm c}$, the long-time relaxation is controlled by the 
diffusive mode.
This relaxation is free of any oscillations and is controlled by the rate 
$s \simeq - 4 \pi^2 A^2/(Q \lambda^2N^2)$. 
This diffusive relaxation exclusively concerns the populations of the system. 
The other nondiffusive modes describe the decoherence 
(as well as the relaxation of the populations beside the part controlled by diffusion
when $2 A > Q \lambda^2$). 
These modes decay at rates ${\rm Re}\, s \simeq -2 Q \lambda^2$, i.e., 
on much shorter time scales than the diffusive mode.
The dynamics of the finite $N$ system is studied in detail elsewhere \cite{Esposito}.


\section{Infinite chain \label{Sec.infinite}}

The spectrum of the infinite chain coupled to its environment can be
obtained from the
spectrum of the finite chain in the infinite-size limit $N\to\infty$.
The wavenumber $q$ becomes a continuous parameter varying in the
first Brillouin zone $-\pi \leq q < \pi$.

For given wavenumber, the diffusive eigenvalue $\mu^{(1)}$ or
$s^{(1)}$ given by Eq. (\ref{spec3aae})
remains isolated. Consequently, we obtain the result that the
dispersion relation of diffusion
is exactly given by the analytical expression
\begin{eqnarray}
s_q
&=& 2 \sqrt{Q^2 \lambda^4 - \left(2A \sin \frac{q}{2}\right)^2}
-2Q\lambda^2 = -Dq^2 + O(q^4) \; . \nonumber\\ \label{diff.infinite}
\end{eqnarray}
The diffusion coefficient
\begin{eqnarray}
D = \frac{A^2}{Q \lambda^2} \; , \label{diff.coeff}
\end{eqnarray}
is proportional to the square of the parameter $A$ of the
tight-binding Hamiltonian
and inversely proportional to the parameter $Q\lambda^2$ of the
coupling to the environment.
The transport is therefore due to the tunneling from site to site,
which is hindered by
the environmental fluctuations proportional to $Q\lambda^2$.
It as been shown in Ref. \cite{Esposito} that, for an Ohmic coupling 
to the environment, the diffusion coefficient is inversely proportional
to the temperature. By using the Einstein relation between the
diffusion coefficient and the conductivity, this latter is therefore
inversely proportional to the square of the temperature.

\begin{figure}[h]
\centering
\begin{tabular}{c@{\hspace{0.5cm}}c}
\vspace*{1cm}
\rotatebox{0}{\scalebox{1.0}{\includegraphics{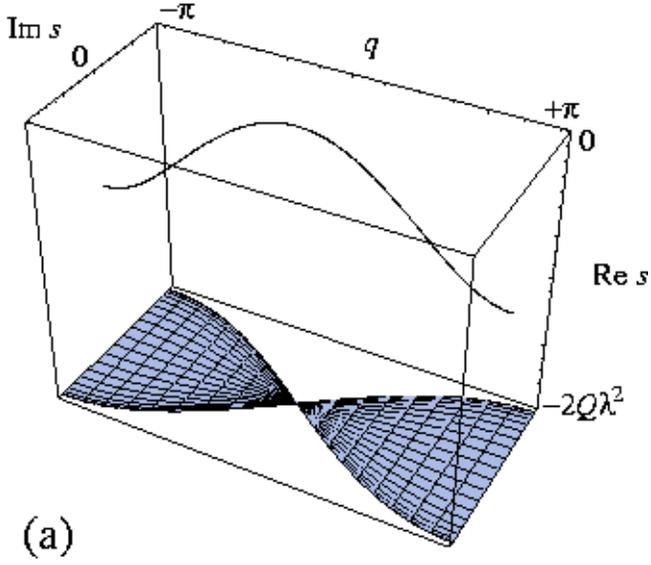}}} \\
\vspace*{1cm}
\rotatebox{0}{\scalebox{1.0}{\includegraphics{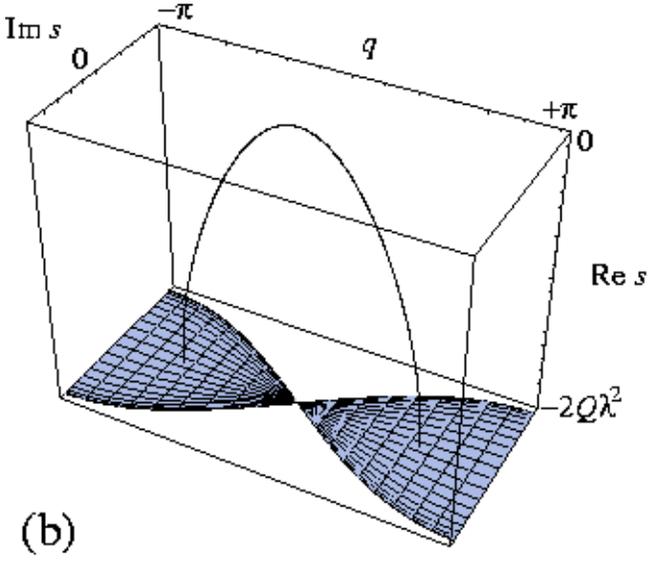}}} \\
\end{tabular}
\caption{Spectrum of the infinite chain coupled to its environment:
(a) in the regime $2A<Q\lambda^2$ for $A=0.4$ and $Q\lambda^2=1$;
(b) in the regime $2A>Q\lambda^2$ for $A=0.6$ and $Q\lambda^2=1$
where the diffusive branch is limited to the wavenumbers $\vert
q\vert < q_{\rm c}=1.97022$.}.
\label{spectre.infinite}
\end{figure}

For $2A \leq Q\lambda^2$, the diffusive eigenvalue exists for all the
values of the wavenumber $-\pi \leq q < +\pi$, as seen in Fig.
\ref{spectre.infinite}(a).

For $2A > Q\lambda^2$, the diffusive eigenvalue only exists for all
the values of the wavenumber in the range
\begin{eqnarray}
-q_{\rm c} \leq q \leq +q_{\rm c} \; , \quad \mbox{with} \quad q_{\rm
c}= 2 \arcsin \frac{Q\lambda^2}{2A} \; .
\label{q_c}
\end{eqnarray}
as seen in Fig. \ref{spectre.infinite}(b).

Beside the isolated diffusive eigenvalue, the spectrum at given
wavenumber $q$ contains a continuous part obtained by the
accumulation of the eigenvalues $\mu^{(2)}$, $\mu^{(3)}$,
$\mu^{(4)}$, and $\mu^{(5)}$ in the limit $N\to\infty$. Indeed, in
this limit, all these eigenvalues accumulate into a
segment of straight line given by
\begin{eqnarray}
s_{q\theta}
&=& -4 i A \left( \sin \frac{q}{2}\right)\; \cos\theta - 2 Q\lambda^2
\; , \label{non.diff.infinite}
\end{eqnarray}
with $0\leq \theta\leq \pi$ and $-\pi \leq q < +\pi$ [see Eqs.
(\ref{spec3abd}) and (\ref{spec3acd})]. This part of the spectrum is
also depicted in Fig. \ref{spectre.infinite} and describes the time
evolution of the quantum coherences which are damped at the
exponential rate $-{\rm Re} \; s_{q\theta} = 2Q\lambda^2$
with possible oscillations due to their nonvanishing imaginary part
${\rm Im} \; s_{q\theta}$.

The eigenvalue (\ref{diff.infinite}) can be considered as a
Liouvillian resonance \cite{Pillet1,Pillet2} similar to the
Pollicott-Ruelle resonances describing diffusion in classical systems
\cite{Gasp96,Gasp98}.


\section{Conclusions \label{Sec.conclusions}}

In the present paper, we have studied an exactly solvable model of
simple translationally invariant subsystems interacting with their
environment. The coupling to the environment is described by
correlation functions which are delta-correlated in space and time.
The reduced dynamics of the subsystem is described by a Redfield
quantum master equation
which takes, for such environments, a Lindblad form.
Thanks to the invariance under spatial translations, we can apply
the Bloch theorem
to the subsystem density matrix. In this way, we succeeded in
getting analytical expressions for all the eigenvalues of the
Redfield superoperator. These eigenvalues control the time evolution
of the subsystem and its relaxation to the thermodynamic equilibrium.
Two kinds of eigenvalues were obtained: the isolated eigenvalue
(\ref{diff.infinite}) giving the dispersion
relation of diffusion along the one-dimensional subsystem and the
other eigenvalues (\ref{non.diff.infinite}) which describe the decay
of the populations and quantum coherences.
The process of decoherence in the subsystem is controlled
by these latter eigenvalues (\ref{non.diff.infinite}).

The properties of the system depend on the length $N$ of the
one-dimensional chain,
on the width $4A$ of the energy band of the
unperturbed tight-binding Hamiltonian and on the
intensity $Q$ of the environmental noise multiplied in the
combination $Q\lambda^2$
with the square of the coupling parameter $\lambda$ of perturbation theory.

We discovered that, for a finite chain, there are two regimes depending on the
the chain length $N$ and the physical parameters $A$ and $Q\lambda^2$.

For a finite and small enough chain, there is a nondiffusive regime
characterized
by a time evolution with oscillations damped by decay rates
proportional to $Q\lambda^2$.
The oscillations are the time evolution of the quantum coherences.
This nondiffusive regime exists if the coupling parameter is smaller
than a critical value which is inversely proportional to the square
root of the chain size $N$:
$\lambda < \lambda_{\rm c}=O(N^{-\frac{1}{2}})$.

For larger chains, we are in the diffusive regime with a monotonic
decay on long times
at a rate controlled by the diffusion coefficient.
In this regime, the slower relaxation mode relaxes exponentially in time
with the scaling $t/(\lambda N)^2$.

In the limit of an infinite chain $N \to \infty$ and for
non-vanishing coupling parameter $Q\lambda^2$,
the nondiffusive regime disappears and the system always diffuses.

The diffusion coefficient is proportional to the square of the width
$4A$ of the energy band
and inversely proportional to the intensity $Q\lambda^2$
of the environmental noise. Accordingly, we are in the presence of a
mechanism of diffusion
in which the quantum tunneling of the particle from site to site
is perturbed by the environmental fluctuations.

The eigenvalues of the Redfield superoperator obtained in
the present paper give the Liouvillian resonances at the second order
of perturbation theory.  In this regard, the present work extends
the results of Refs. \cite{Pillet1,Pillet2} on the spin-boson model to systems
with a translational symmetry in space and capable of sustaining 
the transport property of diffusion beside simple decay processes.
These Liouvillian resonances are the quantum analogues of the 
Pollicott-Ruelle resonances which have been studied elsewhere
for diffusion in classical systems \cite{Gasp96,Gasp98}.


\begin{acknowledgments}
The authors thank Professor G. Nicolis for support and
encouragement in this research. M.~E. is
supported by the ``Fond pour la formation \`{a} la Recherche dans
l'Industrie et dans l'Agriculture". This research is financially supported by the
``Communaut\'e fran\c caise de Belgique"
(``Actions de Recherche Concert\'ees", contract No. 04/09-312),
the National Fund for Scientific Research (F.~N.~R.~S. Belgium), 
the F. R. F. C. (contracts Nos. 2.4542.02 and 2.4577.04), and the U.L.B..
\end{acknowledgments}

\begin{table*}
\caption{{\bf For wavenumbers $q$ corresponding to $\beta <
\frac{1}{2}$:} List of the eigenvalues $\mu_{q\theta}$ of the matrix
(\ref{spec1aae}) given by Eq. (\ref{spec3abd}). These eigenvalues are
directly related to the Redfield superoperator eigenvalues by Eq.
(\ref{spec3acd}).}
\begin{ruledtabular}
\begin{tabular}{llll}
$N$ odd &
$N=4I$ and $q$ even &
$N=4I$ and $q$ odd
\\
&
or $N=4I+2$ and $q$ odd &
or $N=4I+2$ and $q$ even
\\
\hline
\hline
\\
$\mu^{(1)}=\sqrt{1-(2 \beta)^2} + 2 \frac{{\rm e}^{- 2 N {\rm
arccosh} \; 1/(2 \beta)}}{\sqrt{1-(2 \beta)^2}}$ &
$\mu^{(1)}=\sqrt{1-(2 \beta)^2} + 2 \frac{{\rm e}^{- N {\rm arccosh}
\; 1/(2 \beta)}}{\sqrt{1-(2 \beta)^2}}$ &
$\mu^{(1)}=\sqrt{1-(2 \beta)^2} - 2 \frac{{\rm e}^{- N {\rm arccosh}
\; 1/(2 \beta)}}{\sqrt{1-(2 \beta)^2}}$
\vspace{0.5cm}
\\
\hline
\\
$\mu^{(2)} \stackrel{\beta \to 0}{=} - 2i \beta \cos \frac{n \pi}{N}$
&
$\mu^{(2)} \stackrel{\beta \to 0}{=} - 2i \beta \cos \frac{(2n+1)
\pi}{N}$ &
$\mu^{(2)} \stackrel{\beta \to 0}{=} - 2i \beta \cos \frac{2n \pi}{N}$
\\
$\hspace{1.0cm} + \frac{4 \beta^2}{N} \sin^2 \frac{n \pi}{N} $ &
$\hspace{1.0cm} + \frac{8 \beta^2}{N} \sin^2 \frac{(2n+1) \pi}{N}$ &
$\hspace{1.0cm} + \frac{8 \beta^2}{N} \sin^2 \frac{2n \pi}{N}$
\\
$\hspace{1.0cm} + \frac{16 \beta^4}{3 N} \sin^4 \frac{n \pi}{N}$
&
$\hspace{1.0cm} + \frac{32 \beta^4}{3 N} \sin^4 \frac{(2n+1) \pi}{N}$
&
$\hspace{1.0cm} + \frac{32 \beta^4}{3 N} \sin^4 \frac{2n \pi}{N}$
\\
$\hspace{1.0cm} + O(\frac{\beta^6}{N})+ O(\frac{i \beta^3}{N^2})$ &
$\hspace{1.0cm} + O(\frac{\beta^6}{N})+ O(\frac{i \beta^3}{N^2})$ &
$\hspace{1.0cm} + O(\frac{\beta^6}{N})+ O(\frac{i \beta^3}{N^2})$
\\
\hspace{1.0cm} for $n=1,2,\hdots,N-1$ &
\hspace{1.0cm} for $n=0,1,\hdots,\frac{N}{2}-1$ &
\hspace{1.0cm} for $n=1,2,\hdots,\frac{N}{2}-1$
\vspace{0.5cm}
\\
\hline
\\
&
$\mu^{(3)}=- 2i \beta \cos(\frac{2n \pi}{N})$ &
$\mu^{(3)}=- 2i \beta \cos(\frac{(2n+1) \pi}{N})$
\\
&
\hspace{0.6cm} for $n=1,2,\hdots,\frac{N}{2}-1$ &
\hspace{0.6cm} for $n=0,1,\hdots,\frac{N}{2}-1$
\vspace{0.5cm}
\end{tabular}
\end{ruledtabular}
\label{table1}
\end{table*}
\begin{table*}
\caption{{\bf For wavenumbers $q$
corresponding to $\beta > \frac{1}{2}$:} List of the eigenvalues
$\mu_{q\theta}$ of the matrix (\ref{spec1aae}) given by Eq.
(\ref{spec3abd}).
These eigenvalues are directly related to the Redfield superoperator
eigenvalues by Eq. (\ref{spec3acd}).}
\begin{ruledtabular}
\begin{tabular}{llll}
$N$ odd &
$N=4I$ and $q$ even &
$N=4I$ and $q$ odd
\\
&
or $N=4I+2$ and $q$ odd &
or $N=4I+2$ and $q$ even
\\
\hline
\hline
\\
\underline{If $\vert \mu^{(4)} \vert < \vert \mu_c \vert$:}\footnotemark[1]
\vspace{0.5cm}
\\
$\mu^{(4)} \stackrel{\beta \to \infty}{=} - 2i \beta \cos
\frac{(n+\frac{1}{2}) \pi}{N}$ &
$\mu^{(4)} \stackrel{\beta \to \infty}{=} - 2i \beta \cos \frac{2 n
\pi}{N}$ &
$\mu^{(4)} \stackrel{\beta \to \infty}{=} - 2i \beta \cos \frac{(2
n+1) \pi}{N}$
\\
$\hspace{1.0cm}+ \frac{1}{N} (1+\frac{1}{12 \beta^2 \sin^2
\frac{(n+(1/2)) \pi}{N}}$ &
$\hspace{1.0cm}+ \frac{2}{N} (1+\frac{1}{12 \beta^2 \sin^2 \frac{2 n
\pi}{N}}$ &
$\hspace{1.0cm}+ \frac{2}{N} (1+\frac{1}{12 \beta^2 \sin^2 \frac{(2
n+1) \pi}{N}}$
\\
$\hspace{2cm}+\frac{1}{32 \beta^4 \sin^4 \frac{(n+(1/2)) \pi}{N}})$ &
$\hspace{2cm}+\frac{1}{32 \beta^4 \sin^4 \frac{2 n \pi}{N}})$ &
$\hspace{2cm}+\frac{1}{32 \beta^4 \sin^4 \frac{(2 n+1) \pi}{N}})$
\\
$\hspace{1.0cm} + O(\frac{1}{\beta^6 N})+ O(\frac{i}{\beta N^2})$ &
$\hspace{1.0cm} + O(\frac{1}{\beta^6 N})+ O(\frac{i}{\beta N^2})$ &
$\hspace{1.0cm} + O(\frac{1}{\beta^6 N})+ O(\frac{i}{\beta N^2})$
\\
\hspace{1.0cm} for $n_{\rm min} < n <n_{\rm max}$ &
\hspace{1.0cm} for $n_{\rm min} < n <n_{\rm max}$ &
\hspace{1.0cm} for $n_{\rm min} < n <n_{\rm max}$
\\
\\

&
$\mu^{(5)} \stackrel{\beta \to \infty}{=} \mp 2i \beta + \frac{1}{N}
+ O(\frac{1}{\beta})$ &
\\
\vspace{0.5cm}
\\
\underline{If $\vert \mu^{(2)} \vert > \vert \mu_c \vert$:}\footnotemark[1]
\vspace{0.5cm}
\\
$\mu^{(2)} \stackrel{\beta \to 0}{=} - 2i \beta \cos \frac{n \pi}{N}$
&
$\mu^{(2)} \stackrel{\beta \to 0}{=} - 2i \beta \cos \frac{(2n+1)
\pi}{N}$ &
$\mu^{(2)} \stackrel{\beta \to 0}{=} - 2i \beta \cos \frac{2n \pi}{N}$
\\
$\hspace{1.0cm} + \frac{4 \beta^2}{N} \sin^2 \frac{n \pi}{N} $ &
$\hspace{1.0cm} + \frac{8 \beta^2}{N} \sin^2 \frac{(2n+1) \pi}{N}$ &
$\hspace{1.0cm} + \frac{8 \beta^2}{N} \sin^2 \frac{2n \pi}{N}$
\\
$\hspace{1.0cm} + \frac{16 \beta^4}{3 N} \sin^4 \frac{n \pi}{N}$
&
$\hspace{1.0cm} + \frac{32 \beta^4}{3 N} \sin^4 \frac{(2n+1) \pi}{N}$
&
$\hspace{1.0cm} + \frac{32 \beta^4}{3 N} \sin^4 \frac{2n \pi}{N}$
\\
$\hspace{1.0cm} + O(\frac{\beta^6}{N})+ O(\frac{i \beta^3}{N^2})$ &
$\hspace{1.0cm} + O(\frac{\beta^6}{N})+ O(\frac{i \beta^3}{N^2})$ &
$\hspace{1.0cm} + O(\frac{\beta^6}{N})+ O(\frac{i \beta^3}{N^2})$
\\
\hspace{1.0cm} for $n=1,2,\hdots,n_{\rm min}$ &
\hspace{1.0cm} for $n=0,1,\hdots,n_{\rm min}$ &
\hspace{1.0cm} for $n=1,2,\hdots,n_{\rm min}$
\\
\hspace{1.0cm} and $n=n_{\rm max},...,N-1$ &
\hspace{1.0cm} and $n=n_{\rm max},...,\frac{N}{2}-1$ &
\hspace{1.0cm} and $n=n_{\rm max},...,\frac{N}{2}-1$
\\
\vspace{0.5cm}
\\
\hline
\\
&
$\mu^{(3)}=- 2i \beta \cos(\frac{2n \pi}{N})$ &
$\mu^{(3)}=- 2i \beta \cos(\frac{(2n+1) \pi}{N})$
\\
&
\hspace{0.6cm} for $n=1,2,\hdots,\frac{N}{2}-1$ &
\hspace{0.6cm} for $n=0,1,\hdots,\frac{N}{2}-1$
\vspace{0.5cm}
\end{tabular}
\end{ruledtabular}
\footnotetext[1]{$\mu_c = \pm i \sqrt{(2 \beta)^2-1}$}
\label{table2}
\end{table*}

\end{document}